\documentclass[twocolumn,aps,prl]{revtex4}
\usepackage{amsmath}
\usepackage{graphicx}

\begin{document}

\title{Heat transfer between elastic solids with randomly rough surfaces}
\author{ B.N.J. Persson$^1$, B. Lorenz$^1$ and A.I. Volokitin$^{1,2}$}
\affiliation{$^1$IFF, FZ-J\"ulich, 52425 J\"ulich, Germany, EU}
\affiliation{$^2$Samara State Technical University, 443100 Samara,
Russia}

\begin{abstract}
We study the heat transfer between elastic solids with randomly
rough surfaces. We include both the heat transfer from the area of
real contact, and the heat transfer between the surfaces in the
non-contact regions. We apply a recently developed contact mechanics
theory, which accounts for the hierarchical nature of the contact
between solids with roughness on many different length scales. For
elastic contact, at the highest (atomic) resolution the area of real
contact typically consists of atomic (nanometer) sized regions, and
we discuss the implications of this for the heat transfer. For
solids with very smooth surfaces, as is typical in many modern
engineering applications, the interfacial separation in the
non-contact regions will be very small, and for this case we show
the importance of the radiative heat transfer associated with the
evanescent electromagnetic waves which exist outside of all bodies.
\end{abstract}

\maketitle

\pagestyle{empty}


\textbf{1. Introduction}

The heat transfer between solids is a topic of great importance. Classical
applications include topics such as cooling of microelectronic devices, spacecraft structures, satellite bolted
joints, nuclear engineering, ball bearings, tires and
heat exchangers. Other potential applications involve microelectromechanical systems (MEMS). Heat
transfer is also of crucial importance in friction and wear processes, e.g., rubber friction
on hard and rough substrates depends crucially
on the temperature increase in the rubber-countersurface asperity contact regions\cite{Flash}.

\begin{figure}[tbp]
\includegraphics[width=0.45\textwidth,angle=0]{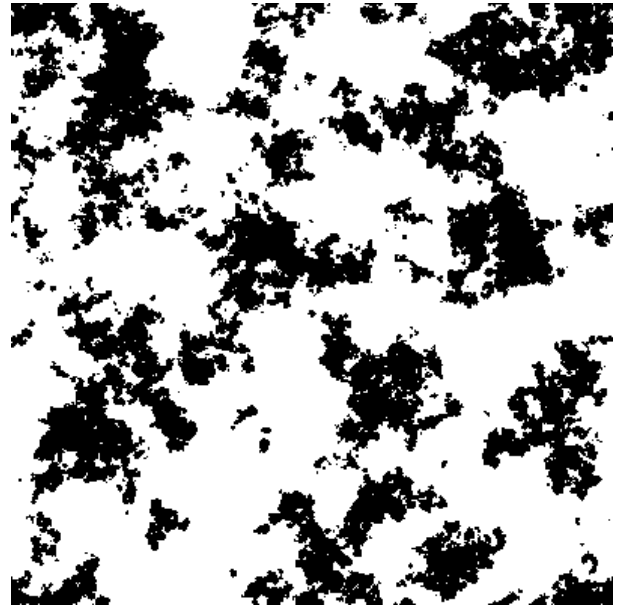}
\caption{
The black area is the contact between two elastic solids with randomly rough surfaces
as obtained using molecular dynamics. For surfaces which have fractal-like roughness
the whole way down to the atomic length scale,
the contact at the highest magnification (atomic resolution) typically consists of nanometer-sized
atomic clusters. Adapted from Ref. \cite{Chunyan1}.}
\label{contact}
\end{figure}

A large number of papers have been published on the heat transfer between randomly rough surfaces\cite{review}.
However, most of these studies are based on asperity contact models such as the model of Greenwood and
Williamson (GW)\cite{GW}. Recent studies have shown that the GW-model (and other asperity contact models\cite{Bush}) are very
inaccurate\cite{inacc1,inacc2}, mainly because of the neglect of the long-range elastic coupling\cite{elast}.
That is, if an asperity is pushed downwards somewhere, the elastic deformation field
extends a long distance away from the asperity, which will
influence the contact involving other asperities further away\cite{Bucher}. This effect is neglected in the GW theory,
but it is included in the contact mechanics model of Persson\cite{JCPpers,PerssonPRL,JCPpers1,PSSR,Chunyan1}, which we use in the present study.
In addition, in the GW model the asperity contact regions are assumed to be circular (or elliptical) while the actual
contact regions (at high enough resolution) have fractal-like boundary lines\cite{Borri,Pei,Chunyan1}, see Fig. \ref{contact}.
Thus, because of their complex nature,
one should try to avoid to directly involve the
nature of the contact regions when studying contact mechanics problems, such as the
heat or electric contact resistance. The approach we use in this paper does not directly involve the nature of the contact regions.
Finally, we note that for elastically hard solids the area of real (atomic) contact $A$
may be a very small fraction of the
nominal or apparent contact area $A_0$, even at high nominal squeezing pressures\cite{P3,BookP}.

\begin{figure}[tbp]
\includegraphics[width=0.45\textwidth,angle=0]{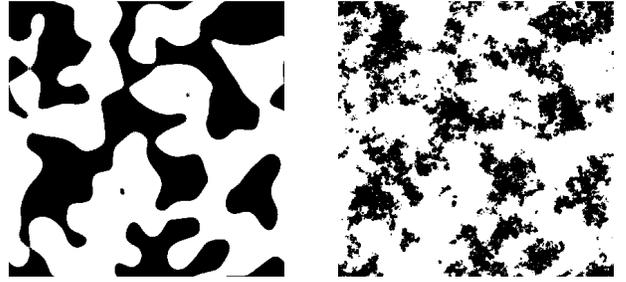}
\caption{
The contact region (black area) between two elastic solids observed at
low (left) and high (right) magnification. The contact resistance
depends mainly on the long-wavelength roughness, and can usually be calculated
accurately from the nature of the contact observed at low magnification (left).}
\label{HeatArea}
\end{figure}

Another important discovery in recent contact mechanics studies is
that for elastic contact, the contact regions observed at atomic
resolution may be just a few atoms wide, i.e., the diameter of the
contact regions may be of the order of $\sim 1 \ {\rm
nm}$\cite{Chunyan,Hyun,Nature1}. The heat transfer via such small
junctions may be very different from the heat transfer through
macroscopic sized contact regions, where the heat transfer usually
is assumed to be proportional to the linear size of the contact
regions (this is also the prediction of the macroscopic heat
diffusion equation), rather than the contact area. In particular, if
the typical phonon wavelength involved in the heat transfer becomes
larger than the linear size of the contact regions (which will
always happen at low enough temperature) the effective heat transfer
may be strongly reduced. Similarly, if the phonons mean free path is
longer than the linear size of the contact regions, ballistic
(phonon) energy transfer may occur which cannot be described by the
macroscopic heat diffusion equation. These effects are likely to be
of crucial importance in many modern applications involving micro
(or nano) sized objects, such as MEMS, where just a few atomic-sized
contact regions may occur. However, for macroscopic solids the
thermal (and electrical) contact resistance is usually very
insensitive to the nature of the contact regions observed at the
highest magnification, corresponding to atomistic (or nanoscale)
length scales. In fact, the heat transfer is determined mainly by
the nature of the contact regions observed at lower magnification
where the contact regions appear larger (see Sec. 5 and
\cite{GreenW,Barber}), see Fig. \ref{HeatArea}. For example, in Sec.
2.2.1 we show that for self-affine fractal surfaces the contact
resistance depends on the range of surface roughness included in the
analysis as $\sim r (H)-(q_0/q_1)^H$, where $q_0$ and $q_1$ are the
smallest and the largest wavevector of the surface roughness
included in the analysis, respectively, and $H$ is the Hurst
exponent related to the fractal dimension via $D_{\rm f} = 3-H$. The
number $r(H)$ depends on $H$ but is of the order of unity. In a
typical case $H\approx 0.8$, and including surface roughness over
one wavevector decade $q_0 < q < q_1 = 10 q_0$ results in a heat
resistance which typically is only $\sim 10\%$ smaller than obtained
when including infinitely many decades of length scales (i.e., with
$q_1 = \infty\times q_0$). At the same time the area of real contact
approaches zero as $q_0/q_1 \rightarrow 0$. Thus, there is in
general no relation between the area of real contact (which is
observed at the highest magnification, and which determines, e.g.,
the friction force in most cases), and the heat (or electrical)
contact resistance between the solids. One aspect of this in the
context of electric conduction was pointed out a long time
ago\cite{Archard}: if an insulating film covers the solids in the
area of real contact, and if electrical contact occurs by a large
number of small breaks in the film, the resistance may be almost as
low as with no film. Similarly, the thermal contact resistance of
macroscopic solids usually does not depend on whether the heat
transfer occur by diffusive or ballistic phonon propagation, but
rather the contact resistance is usually determined mainly by the
nature of the contact regions observed at relative low
magnification.

Note that as $H$ decreases towards zero (or the fractal dimension $D_{\rm f} \rightarrow 3$)
one needs to include more and more decades in the length scales in order to obtain the correct
(or converged) contact resistance, and for $H=0$ (or $D_{\rm f} = 3$) it is necessary to include
the roughness on the
whole way down to the atomic length scale (assuming that the surfaces remain fractal-like with $H=0$
the whole way down to the atomic length scale). Most natural surfaces and surfaces of engineering
interest have (if self-affine fractal) $H > 0.5$ (or $D_{\rm f} < 2.5$), e.g., surfaces prepared
by crack propagation or sand blasting typically have $H\approx 0.8$,  and in these cases the contact
resistance can be calculated accurately from the (apparent) contact observed at relatively low magnification. However,
some surfaces may have smaller Hurst exponents.
One interesting case is surfaces (of glassy solids)
with frozen capillary waves\cite{PSSR,Pires} (which are of great engineering
importance\cite{Pires}), which have $H=0$.
The heat transfer between such surfaces may be understood only by studying the system at the highest
magnification corresponding to atomic resolution.

In this paper we will consider the heat transfer between (macroscopic-sized)
solids in the light of recent advances in contact mechanics.
We will study the contribution to the heat transfer not just from the area of real contact (observed at atomic resolution),
but also the heat transfer across the area of non-contact, in particular the contribution from the fluctuating
electromagnetic field, which surrounds all solid objects\cite{rev1,rev2}. For high-resistivity materials and for hard and
very flat surfaces, such as those involved in many modern applications, e.g., MEMS applications,
this non-contact radiative heat
transfer may in fact dominate in the total heat transfer (at least under vacuum condition).
We note that for flat surfaces (in vacuum)
separated by a distance $d$ larger than the thermal length $d_{\rm T}= c\hbar /k_{\rm B}T$,
the non-contact heat transfer is given by the classical Stefan-Boltzman law, and is independent
of $d$. However, for very short distances the contribution from the evanescent electromagnetic waves to the heat transfer
will be many orders of magnitude larger than the contribution
from propagating electromagnetic waves (as given by the Stefan-Boltzman law)\cite{rev1}.

In most applications (but not in spacecraft applications) one is interested in the heat transfer between solid objects
located in the normal atmosphere and sometimes in a fluid. Most solid objects in the normal atmosphere have
organic and water contamination layers, which may influence the heat transfer for at least two reasons: (a) Thin (nanometer)
contamination layers may occur at the interface in the asperity contact regions, which will
effect the acoustic impedance of the contact junctions, and hence
the propagation of phonon's between the solids (which usually is the origin of the heat transfer, at least
for most non-metallic systems). (b) In addition, capillary bridges may form in the asperity contact regions and effectively increase
the size of the contact regions and increase the heat transfer.
In the normal atmosphere heat can also be transferred between the
non-contact regions via heat diffusion or (at short separation) ballistic processes in the surrounding gas. For larger
separations convective processes may also be important.

In the discussion above we have assumed that the solids deform elastically and we have neglected the
adhesional interaction between the solids.
The contact mechanics theory of Persson can also be applied to cases where adhesion and plastic flow are important,
and we will briefly study how this may affect the heat transfer.
Most solids have modified surface properties, e.g., metals are usually covered
by thin oxide layers with very different conductivities than the underlying bulk materials. However,
as mentioned above, this may not have any major influence on the contact resistance.

Recently, intense research has focused on heat transfer through atomic or molecular-sized junctions\cite{junction1,junction2}. In light of the discussion
presented above, this topic may also be important for the heat transfer between solids, because of the nanometer-sized
nature of the contact regions between solids with random roughness.

This paper is organized as follows: In Sec. 2 we describe the theory for heat transfer between two solids with randomly rough surfaces.
We consider both the heat flow in the area of real contact, and between the surfaces in the non-contact area.
Sec. 3 presents a short review of the contact mechanics theory which is
used to obtain the quantities (related to the surface roughness) which determine the heat transfer
coefficient. In Sec. 4 we present numerical results. In Sec. 5 we discuss the influence of plastic flow and adhesion
on the heat transfer. Sec. 6 presents an application to the heat transfer between tires and the air and road surface. In Sec. 7 we
discuss a new experiment. In Sec. 8 we present experimental results. In Sec. 9 we point out that the developed theory can also be
applied to the electric contact resistance. Sec. 10 contains the summary and conclusion. Appendix A-E present details
related to the theory development and some other general information relevant to the present study.

\begin{figure}[tbp]
\includegraphics[width=0.4\textwidth,angle=0]{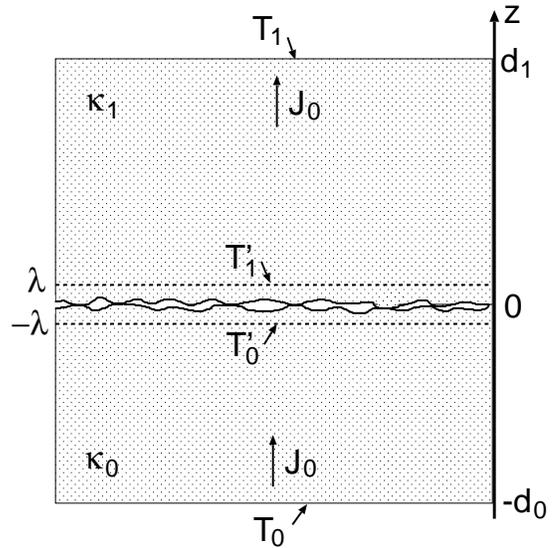}
\caption{
Two elastic solids with nominally flat surfaces squeezed together with the nominal
pressure $p_0$.
The heat current $J_{\rm z}({\bf x})$ at the contacting interface varies strongly with
the coordinate ${\bf x} = (x,y)$ in the $xy$-plane. The average heat current is denoted
by $J_0= \langle J_{\rm z}({\bf x})\rangle$.}
\label{contactblock}
\end{figure}

\vskip 0.3cm \textbf{2. Theory}

\vskip 0.1cm \textbf{2.1 Heat transfer coefficient}

Consider two elastic solids (rectangular blocks) with randomly rough surfaces
squeezed in contact as illustrated in Fig. \ref{contactblock}.
Assume that the temperature at the outer surfaces $z=-d_0$ and $z=d_1$
is kept fixed at $T_0$ and $T_1$, respectively, with $T_0 > T_1$.
Close to the interface
the heat current will vary rapidly in space, ${\bf J} = {\bf J} ({\bf x},z)$, where ${\bf x}=(x,y)$ denote the
lateral coordinate in the $xy$-plane. Far from the interface we will assume that the heat current is constant
and in the $z$-direction, i.e., ${\bf J}= J_0\hat z$.
We denote the average distance between the macro asperity contact regions by $\lambda$ (see Ref. \cite{PSSR}). We assume that
$\lambda << L$, where $L$ is the linear size of the apparent contact
between the elastic blocks. The temperature a distance $\sim \lambda$ from the contacting interface will
be approximately independent of the lateral coordinate ${\bf x} = (x,y)$ and we denote this temperature by
$T_0'$ and $T_1'$ for $z= -\lambda$ and $z=\lambda$, respectively. The heat current for $|z| >> \lambda$
is independent of ${\bf x}$ and can be written as (to zero order in $\lambda /d_0$ and $\lambda /d_1$):
$$J_0=-\kappa_0 {T_0'-T_0\over d_0} = -\kappa_1 {T_1-T_1'\over d_1},\eqno(1)$$
where $\kappa_0$ and $\kappa_1$ are the heat conductivities of the two solid blocks.
We assume that the heat transfer across the interface is proportional to
$T_0'-T_1'$ and we define the heat transfer coefficient $\alpha$ so that
$$J_0=\alpha (T_0'-T_1')\eqno(2)$$
Combining (1) and (2) gives
$$J_0={T_0-T_1\over d_0 \kappa_0^{-1} +d_1 \kappa_1^{-1}+\alpha^{-1}}\eqno(3)$$
This equation is valid as long as $\lambda << L$ and $\lambda << d_0, \ d_1$.
Note that $\alpha$ depends on the macroscopic (or nominal) pressure which act at the interface.
Thus if the macroscopic pressure is non-uniform, as is the case in many
practical applications, e.g., when a ball is squeezed against a flat, one need to include the
dependence of $\alpha$ on ${\bf x}$. Thus in general
$$J({\bf x}) = \alpha ({\bf x}) \left [ T_0'({\bf x})-T_1'({\bf x})\right ]\eqno(4)$$

One expect the contribution to $\alpha$ from the area of real contact to be proportional to
the heat conductivity $\kappa$ (for simplicity we assume here two solids of the same material).
Assuming only elastic deformation, contact mechanics theories show that
for low enough squeezing pressure $p_0$, the area of real contact is proportional to $p_0$,
and the size distribution of contact regions (and the interfacial stress probability distribution) are independent
of $p_0$. Thus one expect that $\alpha$ is proportional to $p_0$.
For randomly rough surfaces
the contact mechanics depends only on the (effective) elastic modulus $E^*$ and
on the surface roughness power spectrum $C(q)$. Thus the only
way to construct a quantity which is proportional to $p_0 \kappa$ and
with the same dimension as $J_0/\Delta T$, using the
quantities which characterize the problem, is
$$\alpha \approx {p_0 \kappa \over E^* u_0}$$
where $u_0$ is a length parameter which is determined from the surface roughness
power spectrum $C(q)$. For self-affine fractal surfaces, $C(q)$ depends only on the root-mean-square roughness
$h_{\rm rms}$, the fractal dimension $D_{\rm f}$ which is dimension less, and on the low and high
cut-off wavevectors $q_0$ and $q_1$. Thus in this case $u_0 = h_{\rm rms} f(D_{\rm f}, q_0/q_1, q_0h_{\rm rms})$.
This result is consistent with the analysis presented in Sec. 2.2.1.
Using the GW-theory result in an expression for $\alpha$ of the form
given above, but with a different function $f$  which now
(even for low squeezing pressures)
also depends
on $p_0/E^*$ (see, e.g., Ref. \cite{Popov}).

\vskip 0.1cm \textbf{2.2 Calculation of $\alpha$}

The heat current ${\bf J}$ and the heat energy density $Q$ are assumed to be given by
$${\bf J} = -\kappa \nabla T, \ \ \ \ \ \ \  Q=\rho C_{\rm V} T$$
where $\kappa$ is the heat conductivity, $\rho$ the mass density and $C_V$ the heat capacitivity.
We consider a steady state condition where $Q$ is time independent. Thus the heat energy continuity equation
$$\nabla \cdot {\bf J} + {\partial Q \over \partial t} =0 $$
reduces to
$$\nabla^2 T = 0$$
We assume that the surface roughness at the interface is so small that when solving the heat flow
equation we can consider the surfaces as flat. However the heat flow across the interface will be highly non-uniform
and given by the heat current $J_z({\bf x})$ (we assume $|\nabla h| << 1$, where $h({\bf x})$ is the surface
height profile). Let us first study the heat flow in the upper solid. We can take into account
the heat flow from the lower solid by introducing a heat source at the interface $z=0$ i.e.
$$\nabla^2 T= -2 J_z ({\bf x})\delta (z)/\kappa_1\eqno(5)$$
Similarly, when studying the temperature in the lower solid we introduce a heat sink on the surface $z=0$ so that
$$\nabla^2 T= 2 J_z({\bf x})\delta (z)/\kappa_0\eqno(6)$$
Let us first study the temperature for $z > 0$. We write
$$J_z({\bf x}) = \int d^2q \ J_z({\bf q}) e^{i {\bf q}\cdot {\bf x}}\eqno(7)$$
$$J_z({\bf q}) = {1\over (2 \pi )^{2}} \int d^2x \ J_z ({\bf x}) e^{-i {\bf q}\cdot {\bf x}}\eqno(8)$$
From (5) we get
$$T({\bf x}, z) = T_1 - {1\over \kappa_1} J_0 (z-d_1)$$
$$ -{1 \over \pi \kappa_1} \int d^2q dk {\Delta J_z({\bf q})\over -q^2-k^2} e^{i({\bf q}\cdot {\bf x}+kz)}\eqno(9)$$
where $J_0 = \langle J_z({\bf x}) \rangle$ is the average heat current and
$$\Delta J_z({\bf x}) = J_z({\bf x})-J_0\eqno(10)$$
Performing the $k$-integral in (9) gives
$$T({\bf x}, z) = T_1 - {1\over \kappa_1} J_0 (z-d_1)$$
$$ +{1\over \kappa_1} \int d^2q \ {1\over q} \Delta J_z ({\bf q}) e^{i{\bf q}\cdot {\bf x}-qz}\eqno(11)$$
Similarly, one obtain for the temperature field for $z<0$:
$$T({\bf x}, z) = T_0 - {1\over \kappa_0} J_0 (z+d_0)$$
$$ - {1\over \kappa_0} \int d^2q \ {1\over q} \Delta J_z({\bf q}) e^{i{\bf q}\cdot {\bf x}+qz}\eqno(12)$$
Let us define
$$\psi({\bf x}) = T({\bf x},-0)-T({\bf x},+0)$$
Using (11) and (12) we get
$$\psi ({\bf x}) = T_0-T_1-\left ({d_0\over \kappa_0}+{d_1\over \kappa_1}\right ) J_0 $$
$$- {1\over \kappa } \int d^2q \ {1\over q}  \Delta J_z({\bf q}) e^{i {\bf q}\cdot {\bf x}}\eqno(13)$$
where
$${1\over \kappa} = {1\over \kappa_0}+{1\over \kappa_1}\eqno(14)$$
From (13) we get
$$\psi({\bf q}) = M \delta ({\bf q}) - {1\over \kappa q} \Delta J_z({\bf q})\eqno(15)$$
where
$$M= T_0-T_1-\left ({d_0\over \kappa_0}+{d_1\over \kappa_1}\right ) J_0\eqno(16)$$

We will now consider two different cases:

\vskip 0.15cm
{\bf 2.2.1 Heat flow through the area of real contact}

Let us consider the area of real contact.
In the contact region $J_z({\bf x})$ will be non-zero but
$\psi({\bf x}) = T({\bf x},+0)-T({\bf x},-0)$ will vanish. On the other surface area
$J_z({\bf x})$ will vanish. Thus we must have
$$J_z({\bf x})\psi ({\bf x}) = 0$$
everywhere. This implies
$$\int d^2q' \ J_z({\bf q}-{\bf q'}) \psi ({\bf q'}) = 0\eqno(17)$$
for all ${\bf q}$.
Combining (15) and (17) gives
$$M J_z({\bf q}) - {1\over \kappa} \int d^2 q' {1\over q'} J_z({\bf q}-{\bf q'}) \Delta J_z({\bf q'})=0$$
The ensemble average of this equation gives
$$M  \langle J_z({\bf q})\rangle - {1\over \kappa} \int d^2 q' {1\over q'}
\langle J_z({\bf q}-{\bf q'}) \Delta J_z({\bf q'})\rangle = 0\eqno(18)$$
From (8) we get
$$\langle J_z({\bf q}=0) \rangle = (2\pi )^{-2} A_0 J_0.$$
Thus the ${\bf q = 0}$ component of (18) gives
$$M A_0 J_0 - {(2\pi )^2 \over \kappa} \int d^2 q {1\over q} \langle |\Delta J_z({\bf q})|^2 \rangle=0\eqno(19)$$
where $A_0$ is the nominal contact area.
Combining (16) and (19) and solving for $J_0$ gives an equation of the form (3) with
$${1\over \alpha} =
{(2\pi )^2 \over \kappa} {1\over A_0 J_0^2}\int d^2 q {1\over q} \langle |\Delta J_z({\bf q})|^2 \rangle \eqno(20)$$

We now assume that the heat current at the interface is proportional to the
normal stress:
$$J_z({\bf x}) \approx \mu \sigma_z ({\bf x}). \eqno(21)$$
We can also write (21) as
$$J_z({\bf x})/J_0  \approx \sigma_z ({\bf x})/p_0,\eqno(22)$$
where $p_0$ is the average pressure. We note that (22) implies that the current density
$J_z({\bf x})$ will be non-vanishing
exactly where the normal stress $\sigma_z ({\bf x})$ is non-vanishing, which must be
obeyed in the present case, where all the heat current flow through the area of real contact.
We note that the heat transfer coefficient depends mainly on the spatial {\it distribution}
of the contact area and this is exactly the same for the pressure
distribution $\sigma ({\bf x})$ as for the
current distribution $J_z({\bf x})$. Thus the fact that in a particular asperity contact region the pressure
$\sigma ({\bf x})$ is not proportional to $J_z({\bf x})$ is not very important in the present context
(see Appendix A and below).

Substituting (22) in (20) gives
$${1\over \alpha} \approx
{(2\pi )^2 \over \kappa} {1\over A_0 p_0^2}\int d^2 q {1\over q} \langle |\Delta \sigma_z({\bf q})|^2 \rangle \eqno(23)$$
We can write
$$\alpha \approx {p_0^2 \kappa \over E^* U_{\rm el}}\eqno(24)$$
where
$$U_{\rm el} =
{(2\pi)^2 \over  A_0 E^*} \int d^2q {1 \over q} \langle |\Delta \sigma ({\bf q})|^2 \rangle\eqno(25)$$
is the stored elastic energy per unit (nominal) surface area\cite{Chunyan1}.
In (25) $E^*$ is the effective elastic modulus
$${1\over E^*} = {1-\nu_0^2\over E_0}+{1-\nu_1^2\over E_1},$$
where $E_0$ and $\nu_0$ are the Young's elastic modulus and the Poisson ratio, respectively, for solid ${\bf 0}$
and similar for solid ${\bf 1}$.
We have shown elsewhere that for small enough load\cite{PerssonPRL}
$U_{\rm el} \approx u_0 p_0$ where $u_0$ is a length of order the root-mean-square surface roughness amplitude.
Thus
$$\alpha \approx {p_0 \kappa \over E^* u_0}.\eqno(26a)$$
Note that for small load the squeezing pressure $p_0$ depends on the (average) interfacial separation $\bar u$ via
the exponential law $p_0 \sim {\rm exp}(-\bar u/u_0)$. Thus the vertical stiffness $dp_0 /d\bar u = - p_0/u_0$ so we can also write
$$\alpha \approx - {\kappa \over E^*} {dp_0 \over d\bar u }.\eqno(26b)$$
This equation is, in fact, exact (see Appendix B and Ref. \cite{Barber}), which shows that the heat transfer is mainly
determined by the geometrical distribution of the contact area (given by the region where
$\sigma_z({\bf x})$ is non-vanishing), and by the thermal interaction between the heat flow through the
various contact spots (see Appendix A).

The length parameter $u_0$ in (26a) can be calculated (approximately)
from the surface roughness power spectrum $C(q)$
using\cite{JCPpers1}
$$u_0 = \surd \pi \int_{q_0}^{q_1} dq \ q^2 C(q) w(q)$$
where
$$w(q) = \left (\pi \int_{q_0}^q dq' q'^3 C(q') \right )^{-1/2}$$
where $q_0$ is the long-distance cut-off (or roll-off) wavevector and $q_1$ the wavevector of the
shortest wavelength roughness included in the analysis.
Assume that the combined surface roughness is self affine fractal for $q_0 < q < q_1$.
In this case
$$C(q) = {H\over \pi} \left ( {h_{\rm rms} \over q_0}\right )^2 \left ({q_0\over q}\right )^{2(H+1)}$$
where $H$ is the Hurst exponent related to the fractal dimension via $D_{\rm f} = 3-H$.
Substituting this $C(q)$ into the equations above gives
$$u_0 \approx \left ({2(1-H)\over \pi H}\right )^{1/2} h_{\rm rms} \left [ r (H)-\left ({q_0\over q_1}\right )^H\right ].$$
where
$$ r(H) = {H\over 2(1-H)}\int_1^\infty dx \ \left (x-1\right )^{-1/2} x^{-1/[2(1-H)]}$$
Note that $ r(H)$ is of order unity (see Ref. \cite{PerssonPRL}).
As discussed in the introduction this implies that the contact resistance in general is determined
accurately by one or two decades of the longest-wavelength roughness components, and that there is no relation between
the area of real contact
(which is observed at the highest magnification, and
which determines, e.g., the friction force in most cases),
and the contact resistance between the solids.

Note that from (3) it follows that one can neglect the heat contact resistance if
$$\kappa /d << \alpha$$
where $\kappa /d $ is the smallest of $\kappa_0 /d_0$ and $\kappa_1 /d_1$. Using (25) this gives
$$d >> u_0 (E^*/p_0)$$
We note that in modern high-tech applications the linear size (or thickness) $d$ of the physical system may be
very small, and in these cases the contact heat resistance may be particular important.

If roughness occurs only on one length scale, say with wavelength $\lambda$ and height $h$, then the pressure
necessary for complete contact will be of order
$$p_0 \approx E^* h/\lambda$$
Substituting this in (26a) gives
$$\alpha \approx \kappa / \lambda\eqno(27)$$
where we have used that $u_0 \approx h$. Thus, $\alpha^{-1} \approx \lambda \kappa^{-1}$ which is the expected result
because the denominator in (3) is only accurate to zero order in $\lambda \kappa^{-1}$. [Alternatively, substituting
(27) in (3) gives a term of the type $(d+\lambda) \kappa^{-1}$ which is the correct
result since $d$ in (3) should really be $d-\lambda$.]

As an example\cite{Bahr}, consider two nominal flat steel plates (in vacuum) with the thickness $d_0=d_1= 0.5 \ {\rm cm}$
and with the root-mean-square roughness $\sim 1 \ {\rm \mu m}$. The plates are squeezed together with
the nominal pressure $p_0 = 0.1 \ {\rm MPa}$.
The ratio between the measured surface and bulk thermal contact resistance is about $150$.
Using (3) we get
$$\Delta T /J_0 = 2d_0 \kappa_0^{-1}+\alpha^{-1}.$$
Thus, the (theoretical) ratio between the surface and the bulk contributions to the thermal resistance is:
$${\kappa_0 \over 2 \alpha d_0},$$
where $\kappa_0$ is the heat conductivity of the bulk steel. Using (25) with $\kappa = \kappa_0/2$ this gives
$${\kappa_0 \over 2 \alpha d_0} = {u_0 \over d_0} {E^*\over p_0}\eqno(28)$$
With (from theory) $u_0 \approx 1 \ {\rm \mu m}$, and $E^* \approx 110 \ {\rm GPa}$, $p_0 = 0.1 \ {\rm MPa}$ and $2d_0= 1 \ {\rm cm}$,
from (28) the ratio between the thermal surface and bulk resistance is $\approx 200$, in good agreement with the experimental data.

The discussion above assumes purely elastic deformations. However, plastic flow is likely to occur
in the present application at short enough length-scales, observed at high magnification.
Since the heat flow is determined mainly by the long-wavelength roughness components,
i.e., by the roughness observed at relative low magnification, when calculating the heat transfer
one may often assume that the surfaces deform purely
elastically, even if plastic deformation is observed at high magnification, see Sec. 5.

\vskip 0.15cm
{\bf 2.2.2 Heat flow through the non-contact area}

Let us now assume that
$$J_z({\bf x}) = \beta ({\bf x}) \left [ T({\bf x},-0)-T({\bf x},+0)\right ]=\beta({\bf x})\psi({\bf x})$$
From (15) we get
$$\psi ({\bf q}) = M \delta ({\bf q})$$
$$- {1\over \kappa q} \int
d^2q' \ \beta({\bf q}-{\bf q'})\left [ 1 - {(2\pi )^2 \over A_0} \delta({\bf q}) \right ] \psi ({\bf q'})\eqno(29)$$
Next, note that
$$J_0={1 \over A_0}\int d^2x \ J_z({\bf x}) = {1\over A_0}\int d^2x \ \beta({\bf x}) \psi ({\bf x})$$
$$={(2\pi )^2 \over A_0 } \int d^2q \ \beta (-{\bf q}) \psi ({\bf q})\eqno(30)$$

Eq. (29) can be solved by iteration. The zero-order solution
$$\psi ({\bf q}) = M  \delta ({\bf q})$$
Substituting this in (30) gives
$$J_0=M {(2\pi )^2 \over A_0} \beta ({\bf q=0})=  M\bar \beta\eqno(31)$$
where
$$\bar \beta = \langle \beta({\bf x}) \rangle = {1\over A_0} \int d^2x \ \beta({\bf x})$$
is the average of $\beta({\bf x})$ over the whole interfacial area $A_0$.
Substituting (16) in (31) and solving for $J_0$ gives an equation of the form (3) with
$\alpha = \bar \beta$.

The first-order solution to (29) is
$$\psi ({\bf q}) = M \delta ({\bf q})-{M\over \kappa q}
\beta({\bf q})\left [ 1 - {(2 \pi)^2 \over A_0} \delta ({\bf q}) \right ]\eqno(32)$$
Substituting (32) in (30) gives again an
equation of the form (3) with
$$\alpha = \bar \beta - {(2\pi)^2 \over \kappa A_0 }\int d^2q {1\over q}
\langle | \beta({\bf q})|^2 \rangle \left [1-{(2 \pi)^2 \over A_0}\delta ({\bf q})\right ], \eqno(33)$$
where we have added $\langle .. \rangle$ which denotes ensemble average, and where we used that
$$\langle\beta({\bf q})\beta(-{\bf q})\rangle = \langle |\beta({\bf q})|^2\rangle$$
We can rewrite (33) as follows.
Let us define the correlation function
$$C_\beta ({\bf q})= {1\over (2\pi)^2}
\int d^2x \ \langle \beta ({\bf x}) \beta ({\bf 0})\rangle e^{i{\bf q}\cdot {\bf x}}\eqno(34)$$
Note that
$$C_\beta ({\bf q})  = {(2\pi )^2 \over A_0} \langle |\beta ({\bf q})|^2\rangle\eqno(35)$$
This equation follows from the fact that the statistical properties are assumed to be translational
invariant in the ${\bf x}$-plane, and is proved as follows:
$$C_\beta ({\bf q})= {1\over (2\pi)^2} \int d^2x \ \langle \beta({\bf x})\beta ({\bf 0}) \rangle e^{i {\bf q}\cdot {\bf x}}$$
$$= {1\over (2\pi)^2} \int d^2x \ \langle \beta({\bf x}+{\bf x'})\beta ({\bf x'}) \rangle e^{i {\bf q}\cdot {\bf x}}$$
$$= {1\over (2\pi)^2} \int d^2x'' \ \langle \beta({\bf x''})\beta ({\bf x'}) \rangle e^{i {\bf q}\cdot ({\bf x''}-{\bf x'})}$$
This equation must be independent of ${\bf x'}$ and we can therefore integrate over the ${\bf x'}$-plane
and divide by the area $A_0$ giving
$$C_\beta ({\bf q}) =
{1\over  (2\pi)^2 A_0}\int d^2x' d^2x'' \ \langle \beta({\bf x''})\beta ({\bf x'})
\rangle e^{i {\bf q}\cdot ({\bf x''}-{\bf x'})}$$
$$= {(2\pi)^2 \over A_0} \langle |\beta ({\bf q})|^2\rangle$$
Let us define
$$\Delta \beta ({\bf x}) =  \beta({\bf x}) - \bar \beta\eqno(36)$$
We get
$$\Delta \beta ({\bf q}) = \beta({\bf q})- \bar \beta \delta ({\bf q})$$
and thus
$$\langle | \Delta \beta ({\bf q})|^2 \rangle = \langle | \beta({\bf q}) |^2  \rangle
\left [1-{(2 \pi)^2 \over A_0}\delta ({\bf q})\right ]\eqno(37)$$
where we have used that
$$\bar \beta \delta({\bf q}) = {(2\pi )^2\over A_0} \beta({\bf q}) \delta ({\bf q})$$
and that
$$\delta ({\bf q}) \delta ({\bf -q}) = \delta ({\bf q}) {1\over (2\pi )^2} \int d^2x \ e^{-i{\bf q}\cdot {\bf x}} = \delta ({\bf q}) { A_0 \over (2\pi )^2}$$
Using (33) and (37) gives
$$\alpha = \bar \beta - {1\over \kappa}\int d^2q q^{-1} C_{\Delta \beta}({\bf q})\eqno(38)$$

\begin{figure}
\includegraphics[width=0.45\textwidth,angle=0]{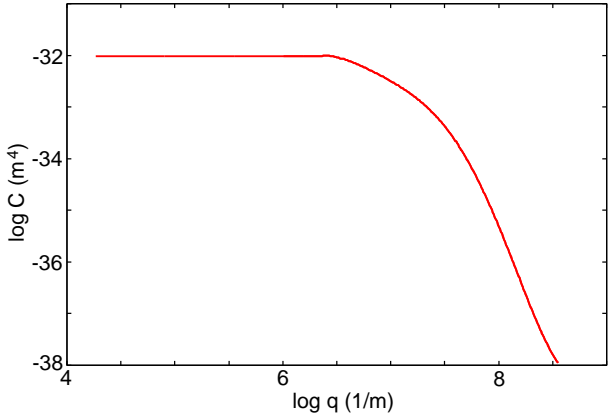}
\caption{\label{PowerSpectrumSiO2}
Surface roughness power spectrum $C(q)$ as a function of the wavevector $q$
on a log-log scale (with 10 as basis). For a typical surface used in MEMS applications
with the root mean square roughness $2.5 \ {\rm nm}$ when measured over an area
$10 \ {\rm \mu m} \times 10 \ {\rm \mu m}$.
}
\end{figure}

\begin{figure}
\includegraphics[width=0.45\textwidth,angle=0]{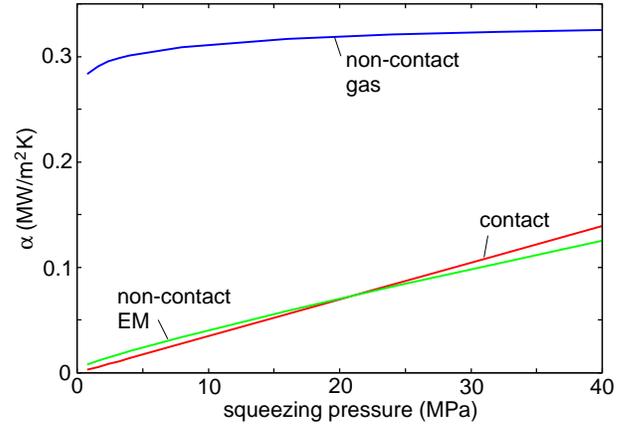}
\caption{\label{HeatAlphaSiO2}
The contribution to the heat transfer coefficient $\alpha$ from
the direct contact area, and the non-contact contribution due to
the fluctuating electromagnetic (EM) field and due to heat transfer
via the surrounding gas.
For
a randomly rough surface with the (combined) surface roughness power spectrum
shown in Fig. \ref{PowerSpectrumSiO2}.
}
\end{figure}

Let us write
$$\langle \Delta \beta ({\bf x}) \Delta \beta ({\bf 0})\rangle  =
\langle (\Delta \beta)^2\rangle f({\bf x})\eqno(39)$$
where $f({\bf 0})=1$. We write
$$f({\bf x}) =  \int d^2q \ f({\bf q}) e^{i{\bf q}\cdot {\bf x}}$$
so that $f({\bf x}={\bf 0}) =1$ gives
$$ \int d^2q \ f({\bf q}) =1\eqno(40)$$
Using (39) and (40), Eq. (38) takes the form
$$\alpha = \bar \beta - \langle (\Delta \beta )^2 \rangle \kappa^{-1} l \eqno(41)$$
where the {\it correlation length}
$$l= {\int d^2 q \ q^{-1} f({\bf q)} \over \int d^2 q \ f({\bf q})}$$
For randomly rough surfaces with isotropic statistical properties $f({\bf q})$ depends only on
$q=|{\bf q}|$ so that
$$l= {\int_0^\infty dq \ f(q) \over \int_0^\infty dq \ q f(q)}$$
Most surfaces of engineering interest are fractal-like, with the surface roughness power spectrum having a
(long-distance) roll-off wavevector $q_0$. In this case one can show from that $l \approx q_0^{-1}$.
For the surface used in the numerical study presented below in Sec. 4 one have $q_0 \approx 10^7 \ {\rm m}^{-1}$
(see Fig. \ref{PowerSpectrumSiO2}). Furthermore, in this case (for amorphous silicon dioxide solids) $\kappa \approx 1 \ {\rm W/mK}$
and if we assume that $\langle (\Delta \beta)^2\rangle $ is of order $ \bar \beta^2$ we get the ratio between the second and the first term in
(41) to be of order $\bar \beta/(q_0 \kappa) \approx 0.01$,
where we have used that typically (see Fig. \ref{HeatAlphaSiO2}) $\bar \beta \approx 0.1 \ {\rm MW /m^2K}$. Thus, in the application presented
in Sec. 4 the second term in the expansion (41) is negligible.

Eq. (41) represent the first two terms in an infinite series which would result if (29) is iterated to infinite order.
The result (41) is only useful if the first term $\bar \beta$ is much larger that the second term. If this is not the case
one would need to include also higher order terms (in principle, to infinite order)
which becomes very hard to calculate using the iterative procedure.
By comparing the magnitude between the two terms in (41) one can determine if it is legitimate to include only the lowest order
term $\bar \beta$.

We now consider two applications of (41), namely the contribution to the heat transfer from
(a) the electromagnetic field (in vacuum) and (b) from heat transfer via a gas (e.g., the normal atmosphere)
which we assume is surrounding the two solids.

\begin{figure}[tbp]
\includegraphics[width=0.45\textwidth,angle=0]{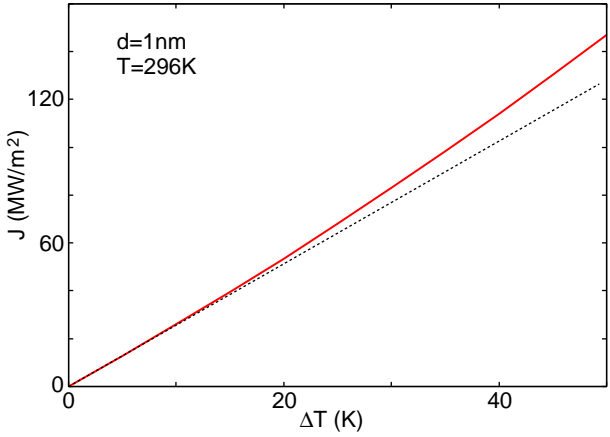}
\caption{
Solid line: The calculated [using (42)] heat current per unit area, $J_0$,  between two (amorphous) silicon dioxide bodies,
as a function of the temperature difference $\Delta T$.
The solids have flat surfaces separated by $d=1 \ {\rm nm}$. One solid is at the temperature
$T=296 \ {\rm K}$ and the other at $T+\Delta T$. Dashed line: linear function with the slope given by the initial slope
(at $\Delta T = 0$) of the solid line.
}
\label{DeltaT.J}
\end{figure}

\vskip 0.1cm \textbf{(a) Radiative contribution to $\alpha$ (in vacuum)}

The heat flux per unit area between two black-bodies separated by $d>> d_T= c\hbar /k_BT$ is given by
the Stefan-Boltzmann law
$$J_0 = {\pi^2 k_{\rm B}^4 \over 60 \hbar^3 c^2} \left (T_0^4-T_1^4\right )$$
where $T_0$ and $T_1$ are the temperatures of solids ${\bf 1}$ and ${\bf 2}$, respectively,
and $c$ the light velocity. In this limiting case the heat transfer between the bodies is determined by
the propagating electromagnetic waves radiated by the bodies and does not depend on the separation
$d$ between the bodies. Electromagnetic waves (or photons) always exist outside any body due to thermal
or quantum fluctuations of the current density inside the body. The electromagnetic field created by
the fluctuating current density exists also in the form of evanescent waves, which are damped exponentially
with the distance away from the surface of the body. For an isolated body, the evanescent waves do not give
a contribution to the energy radiation. However, for two solids separated by $d < d_{T}$, the heat transfer may
increase by many orders of magnitude due to the evanescent electromagnetic waves--this is often referred to
as photon tunneling.

For short separation between two solids with flat surfaces ($d << d_{T}$), the heat current due to the
evanescent electromagnetic waves is given by\cite{rev1}
$$J_0 = {4\over (2\pi)^3} \int_0^\infty d\omega \ \left (\Pi_0(\omega)-\Pi_1(\omega)\right )$$
$$\times \int d^2q \ e^{-2qd} {{\rm Im} R_0(\omega) {\rm Im} R_1(\omega) \over
|1-e^{-2qd} R_0(\omega)R_1(\omega) |^2}\eqno(42)$$
where
$$\Pi (\omega) = \hbar \omega \left (e^{\hbar \omega /k_{\rm B}T}-1\right )^{-1}$$
and
$$R(\omega) = {\epsilon (\omega) -1 \over \epsilon (\omega) + 1}$$
where $\epsilon (\omega)$ is the dielectric function.
From (42)  it follows that the heat current scale as $1/d^2$ with the separation between the solid surfaces.
The heat current is especially large in the case of resonant photon tunneling between surface modes
localized on the two different surfaces. The resonant condition corresponds
to the case when the denominator in the  integrand of (42) is small. Close to the resonance we
can use the approximation
$$R \approx \frac{\omega_1}{\omega -\omega _0-i\gamma },$$
where $\omega_1$ is a constant and $\omega_0$ is determined by the equation
${\rm Re} [ \epsilon (\omega_0) + 1] =0$.
In this case the heat current is determined
by\cite{rev1}
$$J_0 \approx \mu {\gamma \over d^2}\left[\Pi_0(\omega_0)-\Pi_1(\omega_0)\right], $$
where $\mu \approx [{\rm log} (2\omega _a/\gamma )]^2/(8\pi)$. If we write $T_1=T_0-\Delta T$ and
assume $\Delta T/T_0 << 1$ we get $J_0=\alpha \Delta T$ with
$$\alpha \approx \mu {k_{\rm B} \gamma \over d^2} {\eta^2 {\rm exp} (\eta) \over [{\rm exp} (\eta) -1 ]^2}\eqno(43)$$
where $\eta=\hbar \omega_0 /k_{\rm B}T_0$.

Resonant photon tunneling enhancement of the heat transfer is
possible for two semiconductor or insulator surfaces which can support
low-frequency surface phonon-polariton modes in the mid-infrared frequency
region. As an example,
consider two clean surfaces of (amorphous) silicon dioxide (SiO$_2$). The optical
properties of this material can be described using an oscillator
model\cite{optical}
$$\epsilon (\omega) = \epsilon_\infty + {a\over \omega_a^2 -\omega^2 -i\omega \gamma_a}+{b\over \omega_b^2 -\omega^2 -i\omega \gamma_b}$$
The frequency dependent term in this expression is due to optical phonon's.
The values for the parameters $\epsilon_\infty$, $(a,\omega_a,\gamma_a)$
and $(b,\omega_b,\gamma_b)$ are given in Ref. \cite{optical}.
In Fig. \ref{DeltaT.J} we show
the calculated heat current per unit area, $J_0$,
as a function of the temperature difference $\Delta T$.
The solids have flat surfaces separated by $d=1 \ {\rm nm}$. One solid is at the temperature
$T=296 \ {\rm K}$ and the other at $T+\Delta T$.
When $\Delta T << T$, the heat transfer depends (nearly) linearly on the temperature
difference $\Delta T$ (see Fig. \ref{DeltaT.J}), and we can define the
heat transfer coefficient $\alpha = J_0/\Delta T$. In the present case (for $d=d_0 =1 \ {\rm nm}$)
$\alpha = \alpha_0 \approx 2\times 10^6 \ {\rm W/m^2 K}$.
If the surfaces are not smooth but if roughness occur so that the separation $d$ varies
with the coordinate ${\bf x}=(x,y)$ we have to first order in the expansion (41):
$$\alpha = \bar \beta = \alpha_0 \langle \left (d_0/d \right )^2 \rangle\eqno(44)$$
where $\langle .. \rangle$ stands for ensemble average, or average over the whole surface area, and where
$\alpha_0$ is the heat transfer between flat surfaces separated by $d=d_0$.

In the preset case the heat transfer is associated with thermally excited optical (surface) phonon's. That is, the electric field of a
thermally excited optical phonon in one solid excites an optical phonon in the other solid, leading to energy
transfer. The excitation transfer occur in both directions but if one solid is hotter than the other,
there will be a net transfer of energy from the hotter to the colder solid. For metals, low-energy excited electron-hole pairs
will also contribute to the energy transfer, but for good metals
the screening of the fluctuating electric field by the conduction electrons leads to very ineffective heat transfer.
However, if the metals are covered with metal oxide layers, and if the separation
between the solids is smaller than the oxide layer thickness, the energy transfer may again be due mainly
to the optical phonon's of the oxide, and the magnitude of the heat current will be similar to what we calculated
above for (amorphous) silicon dioxide.

Let us consider a high-tech application. Consider a MEMS device involving very smooth (amorphous) silicon dioxide slabs.
Consider, for example,
a very thin silicon dioxide slab rotating on a silicon dioxide substrate.
During operation a large amount of frictional energy may be generated at the interface.
Assume that the  disk is
pressed against the substrate with the nominal stress or pressure $p_0$. This does not need to be an external
applied force but may be due to the long-ranged van der Waals attraction between the solids, or due to capillary bridges formed in the vicinity of the
(asperity) contact regions between the solids. The heat transfer due to the area of real contact (assuming purely elastic deformation) can
be calculated from (25). Let us make a very rough estimate: Surfaces used in MEMS application have typically a roughness of order a few nanometers.
Thus, $u_0 \sim 1 \ {\rm nm}$ and for (amorphous) silicon dioxide the heat conductivity
$\kappa \approx 1 \ {\rm W/Km}$. Thus from (32):
$$\alpha \approx (p_0/E) \times 10^9 \ {\rm W/m^2 K}\eqno(45)$$
In a typical case the nominal pressure $p_0$ may be (due to the van der Waals interaction and capillary bridges) between $10^6-10^7 \ {\rm Pa}$
and with $E\approx 10^{11} \ {\rm Pa}$ we get from (45) $\alpha \approx 10^4 - 10^5 \ {\rm W/Km^2}$. If the root-mean-square roughness is of order
$\sim 1 \ {\rm nm}$ we expect the average separation between the surfaces to be of order a few nanometer so that $\langle (d_0/d)^2 \rangle \approx 0.1$ giving
the non-contact contribution to $\alpha$ from the electromagnetic field of order [from (44)] $10^5 \ {\rm W/Km^2}$, i.e., larger than
or of similar magnitude as the contribution from the area of real contact.

\vskip 0.1cm \textbf{(b) Contribution to $\alpha$ from heat transfer via the surrounding gas or liquid}

Consider two solids with flat surfaces separated by a distance $d$. Assume that the solids are surrounded by a gas.
Let $\Lambda$ be the gas mean free path. If $d >> \Lambda$ the heat transfer between the solids occurs via
heat diffusion in the gas. If $d << \Lambda$ the heat transfer occurs by ballistic propagation of gas molecules
from one surface to the other. In this case gas molecules reflected from the hotter surface will have (on the average)
higher kinetic energy that the gas molecules reflected from the colder surface. This will result in heat transfer from
the hotter to the colder surface. The heat current is approximately given by\cite{Bahrami1}
$$J_0 \approx {\kappa_{\rm gas} \Delta T \over d + a \Lambda}$$
where $a$ is a number of order unity and which depend on the interaction
between the gas molecules and the solid walls\cite{review}.
For air (and most other gases) at the normal atmospheric pressure and at room temperature $\Lambda \approx 65 \ {\rm nm}$
and $\kappa_{\rm gas} \approx 0.02 \ {\rm W/mK}$.
For contacting surfaces with surface roughness we get to first order in the expansion in (41):
$$\alpha \approx \kappa_{\rm gas} \langle (d+\Lambda )^{-1} \rangle \eqno(46)$$
where $\langle .. \rangle$ stand for ensemble average or averaging over the surface area. Eq. (46) also holds if
the surfaces are surrounded by a liquid rather than a gas. In this case $\kappa_{\rm gas}$ must be replaced with the
liquid heat conductivity $\kappa_{\rm liq}$ and in most cases one can put $\Lambda$ equal to zero.

If we again consider a MEMS application where the average
surface separation is of order nm we can neglect the $d$-dependence in (46) and get
$\alpha \approx \kappa_{\rm gas}/\Lambda \approx 3\times 10^5 \ {\rm W/m^2 K}$ which is similar to the contribution
from the electromagnetic coupling.

\vskip 0.1cm \textbf{(c) Contribution to $\alpha$ from heat transfer via capillary bridges}

If the solid walls are wet by water, in a humid atmosphere capillary bridges will form spontaneous
at the interface in the vicinity of the asperity contact regions. For very smooth surfaces, such as in MEMS
applications, the fluid (in this case water) may occupy a large region between the surfaces and will then
dominate the heat transfer between the solids. Similarly, contamination layers (mainly organic molecules)
which cover most natural surfaces may form capillary bridges between the contacting solids, and
contribute in an important way to the heat transfer coefficient. The fraction of the interfacial
surface area occupied by fluid bridges, and the separation between the solids in the fluid
covered region, can be calculated using the theory developed in Ref. \cite{PerssonCapillary}. From this one can calculate the
contribution to the heat transfer using (46):
$$\alpha \approx \kappa_{\rm liq} \langle d^{-1} \rangle \approx \kappa_{\rm liq} \int_a^{d_{\rm K}} du A_0 P(u) u^{-1}\eqno(47)$$
where $P(u)$ is the distribution of interfacial separation $u$, and $A_0$ the nominal contact area. The lower cut-off
$a$ in the integral is a distance of order a molecular length and $d_{\rm K}$ is the maximum height of the
liquid bridge which, for a system in thermal equilibrium and for a wetting liquid, is of order the Kelvin
length. Note that $P(u)$ is normalized and that
$$\int_a^{d_{\rm K}} du A_0 P(u) = \Delta A\eqno(48)$$
is the surface area (projected on the $xy$-plane) where the surface separation is between $a < u < d_{\rm K}$.

\vskip 0.3cm \textbf{3. Contact mechanics: short review and basic equations}

The theory of heat transfer presented above depends on quantities which can be calculated using contact mechanics theories.
Thus, the heat flux through the non-contact area (Sec. 2.2.2) depends on the average of some function $f[d({\bf x})]$ of
the interfacial separation $d({\bf x})$. If $P(u)$ denote the probability distribution of interfacial separation $u$ then
$$\langle f(d) \rangle = \int_a^\infty du \ f(u) P(u)\eqno(49)$$
where $a$ is a short-distance cut-off (typically of molecular dimension). The contribution from the area of real contact
depends on the elastic energy $U_{\rm el}$ stored in the asperity contact regions [see Eq. (23)]. In the limit of small
contact pressure $U_{\rm el}=p_0 u_0$, where $u_0$ is a length which is of order the root-mean-square
roughness of the combined roughness profile. All the quantities $P(u)$, $U_{\rm el}$ and $u_0$ can be calculated
with good accuracy using the contact mechanics model of Persson. Here we will briefly review this theory
and give the basic equations relevant for heat transfer.

\begin{figure}
\includegraphics[width=0.45\textwidth,angle=0]{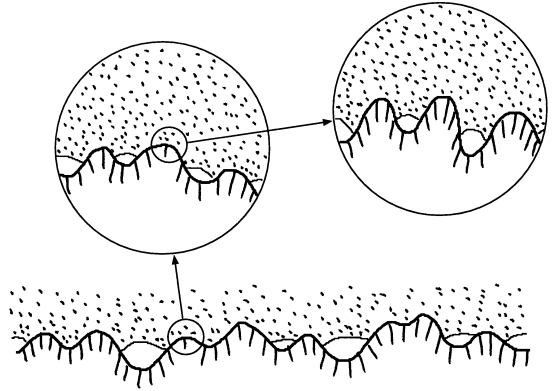}
\caption{\label{1x}
An rubber block (dotted area) in adhesive contact with a hard
rough substrate (dashed area). The substrate has roughness on many different
length scales and the rubber makes partial contact with the substrate on all length scales.
When a contact area
is studied at low magnification it appears as if complete contact occur,
but when the magnification is increased it is observed that in reality only partial
contact occur.
}
\end{figure}

Consider the frictionless
contact between two elastic solids with the Young's elastic modulus $E_0$ and $E_1$ and the Poisson ratios $\nu_0$ and $\nu_1$.
Assume that the solid surfaces have the height profiles $h_0 ({\bf x})$ and $h_1({\bf x})$, respectively. The elastic
contact mechanics for the solids is equivalent to those of a rigid substrate with the height profile $h({\bf x}) = h_0({\bf x})+
h_1({\bf x})$ and a second elastic solid with a flat surface and with the Young's modulus $E$ and
the Poisson ratio $\nu$ chosen so
that\cite{Johnson2}
$${1-\nu^2\over E} = {1-\nu_0^2\over E_0}+{1-\nu_1^2\over E_1}.\eqno(50)$$

\begin{figure}
\includegraphics[width=0.35\textwidth,angle=0.0]{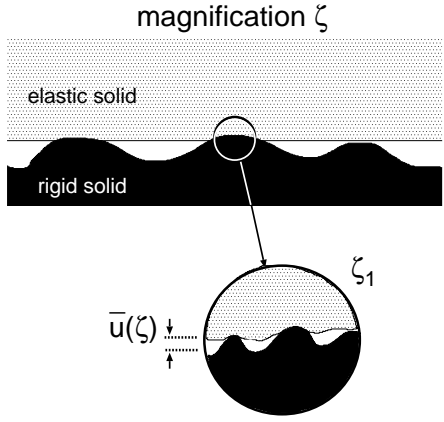}
\caption{\label{asperity.mag}
An asperity contact region observed at the magnification $\zeta$. It appears that
complete contact occur in the asperity contact region, but when the magnification is
increasing to the highest (atomic scale) magnification $\zeta_1$,
it is observed that the solids are actually separated by the average distance $\bar{u}(\zeta)$.
}
\end{figure}

The contact mechanics formalism developed elsewhere\cite{PSSR,JCPpers,PerssonPRL,JCPpers1} is
based on the studying the
interface between two contacting solids at different magnification $\zeta$.
When the system is studied at the magnification $\zeta$ it appears as if the contact area
(projected on the $xy$-plane) equals $A(\zeta)$, but when the magnification
increases it is observed that the contact is incomplete, and the surfaces in the apparent
contact area $A(\zeta)$ are in fact separated by
the average distance $\bar u(\zeta)$, see Fig. \ref{asperity.mag}.
The (apparent) relative contact area $A(\zeta)/A_0$ at the magnification $\zeta$
is given by\cite{JCPpers,JCPpers1}
$${A(\zeta)\over A_0} = {1\over (\pi G )^{1/2}}\int_0^{p_0} d\sigma \ {\rm e}^{-\sigma^2/4G}
= {\rm erf} \left ( p_0 \over 2 G^{1/2} \right )\eqno(51)$$
where
$$G(\zeta) = {\pi \over 4}\left ({E\over 1-\nu^2}\right )^2 \int_{q_0}^{\zeta q_0} dq q^3 C(q)\eqno(52)$$
where the surface roughness power spectrum
$$C(q) = {1\over (2\pi)^2} \int d^2x \ \langle h({\bf x})h({\bf 0})\rangle {\rm e}^{-i{\bf q}\cdot {\bf x}}\eqno(53)$$
where $\langle ... \rangle$ stands for ensemble average.
The height profile $h({\bf x})$ of the rough surface can be measured routinely
today on all relevant length scales using optical and stylus experiments.

We define
$u_1(\zeta)$ to be the (average) height separating the surfaces which appear to come into
contact when the magnification decreases from $\zeta$ to $\zeta-\Delta \zeta$, where $\Delta \zeta$
is a small (infinitesimal) change in the magnification. $u_1(\zeta)$ is a monotonically decreasing
function of $\zeta$, and can be calculated from the average interfacial separation
$\bar u(\zeta)$ and $A(\zeta)$ using
(see Ref.~\cite{JCPpers1})
$$u_1(\zeta)=\bar u(\zeta)+\bar u'(\zeta) A(\zeta)/A'(\zeta),\eqno(54)$$
where\cite{JCPpers1}
$$\bar{u}(\zeta ) = \surd \pi \int_{\zeta q_0}^{q_1} dq \ q^2C(q) w(q)$$
$$\times \int_{p(\zeta)}^\infty dp'
 \ {1 \over p'} e^{-[w(q,\zeta) p'/E^*]^2},\eqno(55)$$
where $E^*=E/(1-\nu^2)$, and
where $p(\zeta)=p_0A_0/A(\zeta)$
and
$$w(q,\zeta)=\left (\pi \int_{\zeta q_0}^q dq' \ q'^3 C(q') \right )^{-1/2}.$$

The distribution of interfacial separations
$$P(u) = \langle \delta [u-u({\bf x})]\rangle$$
where $u({\bf x}) = d({\bf x})$ is the separation between the surfaces at point ${\bf x}$.
As shown in Ref. \cite{JCPpers1} we have (approximately)
$$P(u)= \int_1^\infty d\zeta \ [-A'(\zeta )] \delta [u-u_1(\zeta)]$$
Thus we can write (49) as
$$\langle f(d) \rangle = \int_1^{\zeta_1} d\zeta \ [-A'(\zeta )] f[u_1(\zeta)]\eqno(56)$$
where $\zeta_1$ is defined by $u_1({\zeta_1})=a$.

\begin{figure}
\includegraphics[width=0.45\textwidth,angle=0]{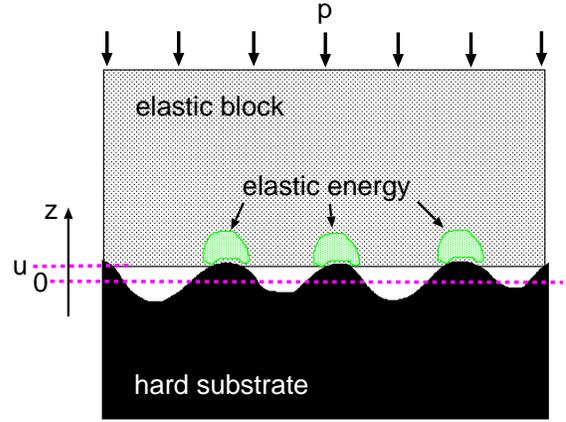}
\caption{\label{block}
An elastic block squeezed against a rigid rough substrate. The separation
between the average plane of the substrate and the average plane of the lower
surface of the block is denoted by $u$. Elastic energy is stored in the block in the vicinity
of the asperity contact regions.
}
\end{figure}

Finally, the elastic energy $U_{\rm el}$ and the length parameter $u_0$ can
be calculated as follows.
The elastic energy $U_{\rm el}$ has been studied in Ref. \cite{elast}:
$$U_{\rm el} = A_0 E^* {\pi \over 2} \int_{q_0}^{q_1} dq \ q^2 W(q,p)C(q).\eqno(57)$$
In the simplest case one take $W(q,p)=P(q,p)=A(\zeta) /A_0$
is the relative contact area when the interface is studied at
the magnification $\zeta = q/q_0$, which depends on the applied pressure $p=p_0$.
A more accurate expression is
$$W(q,p) = P(q,p) \left [\gamma +(1-\gamma) P^2(q,p)\right ].\eqno(58)$$
However, in this case one also need to modify (55) appropriately (see Ref. \cite{JCPpers1}).
The parameter $\gamma$ in (58) seams to depend on the surface roughness. For self-affine fractal surfaces with the fractal
dimension
$D_{\rm f} \approx 2.2$ we have found that $\gamma \approx 0.5$ gives good agreement between the theory and
numerical studies\cite{Chunyan1}.
As $D_{\rm f} \rightarrow 2$ analysis of numerical data indicate that
$\gamma \rightarrow 1$.

For small pressures one can show that\cite{JCPpers1}:
$$p=\beta E^* e^{- \bar u/u_0},\eqno(59)$$
where
$$u_0 = \surd \pi \gamma \int_{q_0}^{q_1} dq \ q^2 C(q) w(q),\eqno(60)$$
where $w(q)=w(q,1)$, and where
$$\beta = \epsilon \
{\rm exp}\left [  {\int_{q_0}^{q_1} dq \ q^2C(q) w(q) {\rm log} w(q)\over
\int_{q_0}^{q_1} dq \ q^2C(q) w(q)}\right ],\eqno(61)$$
where (for $\gamma = 1$) $\epsilon=0.7493$.

\begin{figure}
\includegraphics[width=0.45\textwidth,angle=0]{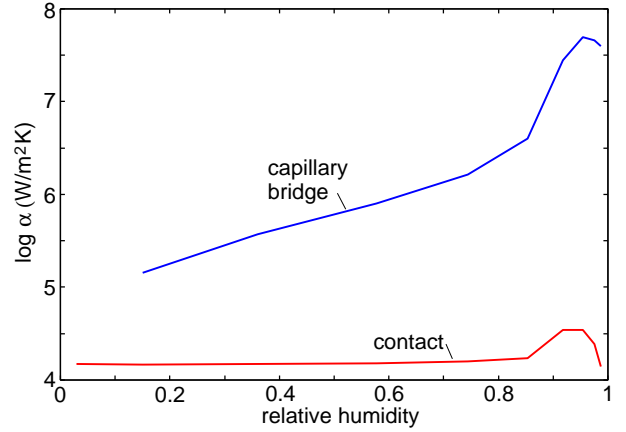}
\caption{\label{relativeHumidity.logAlpha.contact.fluid}
The logarithm (with 10 as basis) of the contribution to the heat transfer coefficient
$\alpha$ from the real contact areas, and from the water in the capillary bridges, as a function
of the relative (water) humidity.
For
a randomly rough surface with the (combined) surface roughness power spectrum
shown in Fig. \ref{PowerSpectrumSiO2}. The squeezing pressure $p_0 = 4 \ {\rm MPa}$ and
the effective solid elastic modulus $E^* = 86 \ {\rm GPa}$.
The heat conductivity of water
$\kappa_{\rm fluid} = 0.58  \ {\rm W/mK}$.
}
\end{figure}

\vskip 0.3cm \textbf{4. Numerical results}

In this section we present numerical results to illustrate the
theory. We focus on a MEMS-like application. In Fig.
\ref{PowerSpectrumSiO2} we show the surface roughness power spectrum
$C(q)$ as a function of the wavevector $q$ on a log-log scale (with
10 as basis) for a typical surface used in MEMS applications, with
the root mean square roughness $2.5 \ {\rm nm}$ when measured over
an area $10 \ {\rm \mu m} \times 10 \ {\rm \mu m}$. In Fig.
\ref{HeatAlphaSiO2} we show for this case the contribution to the
heat transfer coefficient $\alpha$ from the direct contact area, and
the non-contact contribution due to the fluctuating electromagnetic
(EM) field and due to heat transfer via the surrounding gas. In the
calculation of the EM-contribution we have used (44) with $\alpha_0
= 2.0 \ {\rm MW/m^2K}$ (and $d_0 = 1 \ {\rm nm}$). For the
contribution from the surrounding gas we have used (46) with
$\kappa_{\rm gas} = 0.024 \ {\rm W/mK}$ and $\Lambda = 65 \ {\rm
nm}$ (and $a=1$). For the contact contribution we used (25) with
$\kappa = 1 \ {\rm W/mK}$. In all calculations we have assumed $E^*
= 86 \ {\rm GPa}$ and that the contact is elastic (no plastic
yielding).

We have also studied the contribution to the heat transfer from capillary bridges which
on hydrophilic surfaces form spontaneous in a humid atmosphere.
The capillary bridges gives an attractive force (to be added to the external squeezing force),
which pulls the solids closer together. We have used the theory presented in Ref. \cite{PerssonCapillary}
to include the influence of capillary bridges on the contact mechanics, and to determine the fraction
of the interface area filled with fluid at any given relative humidity.
In Fig. \ref{relativeHumidity.logAlpha.contact.fluid} we show
the logarithm (with 10 as basis) of the contribution to the heat transfer coefficient
$\alpha$ from the real contact areas, and from the water in the capillary bridges, as a function
of the relative (water) humidity. For relative humidity below $\sim 0.4$ the contribution to the heat transfer from capillary bridges
decreases roughly linearly with decreasing humidity (and vanish at zero humidity),
and for relative humidity below $\sim 0.015$ the heat transfer
via the area of real contact will be more important than the contribution from the capillary bridges.
However the contribution from heat transfer
via the air or vapor phase (not shown) is about $\sim 0.3 \ {\rm MW/m^2 K}$ (see Fig. \ref{HeatAlphaSiO2}),
and will hence give the dominant contribution to the heat transfer for relative humidity below $0.3$.
The small increase in the contribution from the area of real contact for relative humidity around
$\sim 0.94$ is due to the increase in the contact area due to the force from the capillary bridges.
For soft elastic solids (such as rubber) this effect is much more important: see Ref. \cite{PerssonCapillary} for a detailed discussion
of this effect, which will also affect (increase) the heat transfer in a drastic way.

We note that heat transfer via capillary bridges has recently been observed in nanoscale
point contact experiments\cite{capillary1}. In this study the authors investigated
the heat transfer mechanisms at a $\sim 100 \ {\rm nm}$ diameter point contact between a sample and a probe tip of a scanning thermal
microscope. They observed heat transfer both due to the surrounding (atmospheric) air, and via capillary bridges.

\begin{figure}
\includegraphics[width=0.45\textwidth,angle=0]{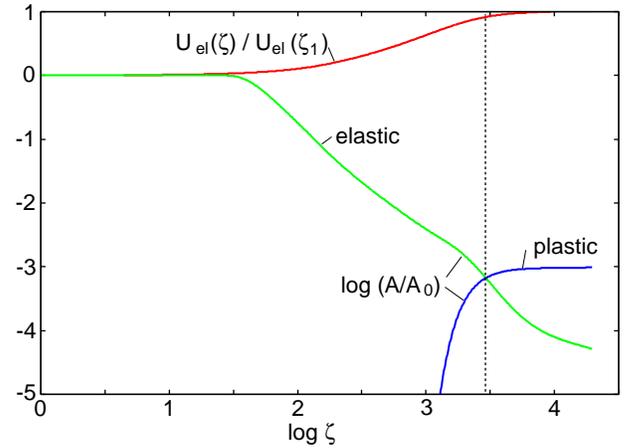}
\caption{\label{log.magnification.u0.logAelast.logAplast}
The elastic $A_{\rm el}$ and plastic $A_{\rm pl}$ contact area as a function of magnification on a
log-log scale (with 10 as basis). The
penetration hardness $\sigma_{\rm Y} = 4 \ {\rm GPa}$ and the applied pressure $p_0= 4 \ {\rm MPa}$.
Also shown is the asperity-induced elastic energy $U_{\rm el}(\zeta)$
in units of the full elastic energy $U_{\rm el} (\zeta_1)$ obtained when
all the roughness (with wavevectors below $q_1 = \zeta_1 q_0$) is included.
The vertical dashed line indicate the magnification where $A_{\rm el} = A_{\rm pl}$.
}
\end{figure}

\vskip 0.3cm \textbf{5. Role of adhesion and plastic deformation}

In the theory above we have assumed that the solids deform purely elastically. However,
in many practical situations the solids will deform plastically at short enough length scale.
Similarly, in many practical situations, in particular for elastically soft solids, the area
of real contact may depend strongly on the adhesive interaction across the contacting interface.
Here we will briefly discuss under which circumstances this will affect the heat transfer
between the solids.

The contribution to the heat transfer from the area of real contact between two solids depends
on the elastic energy $U_{\rm el}$ stored in the asperity contact regions,
or, at small enough applied loads, on the length parameter $u_0$. For most randomly
rough surfaces these quantities are determined mainly by the long-wavelength,
large amplitude surface roughness components. Similarly, the interfacial
separation, which determines the non-contact contribution to the heat transfer,
depends mainly on the long-wavelength, large amplitude surface roughness components.
On the other hand, plastic deformation and adhesion often manifest themself only at
short length scales, corresponding to high magnification. For this reason, in many cases
one may assume purely elastic deformation when calculating the heat transfer, even if,
at short enough length scale, all asperities have yielded plastically, or the adhesion
has strongly increased the (apparent) contact area. Let us illustrate this
with the amorphous silicon dioxide system studied in Sec. 4.

In Fig. \ref{log.magnification.u0.logAelast.logAplast} we show
the elastic and plastic contact area as a function of magnification on a
log-log scale (with 10 as basis).
Also shown is the asperity-induced elastic energy $U_{\rm el}(\zeta)$
in units of the full elastic energy $U_{\rm el}(\zeta_1)$ obtained when
all the roughness (with wavevectors below $q_1 = \zeta_1 q_0$) is included.
Note that about $90 \%$ of the full elastic energy is already obtained at the magnification where
the elastic and plastic contact areas are equal, and about $60 \%$ of the full elastic energy
is obtained when $A_{\rm pl} /A_{\rm el} \approx 0.01$. Thus, in the present case,
to a good approximation, we can neglect the
plastic deformation when studying the heat transfer.
In the calculation we have assumed the penetration hardness $\sigma_{\rm Y} = 4 \ {\rm GPa}$
and the squeezing pressure $p_0 = 4 \ {\rm MPa}$. Thus, at high magnification, where
all the contact regions are plastically deformed, the relative contact area $A/A_0 = p_0/\sigma_{\rm Y} = 0.001$
in good agreement with the numerical data in Fig. \ref{log.magnification.u0.logAelast.logAplast}.

If necessary, it is easy to include adhesion and plastic deformation when calculating the heat transfer
coefficient $\alpha$. Thus (26b) is also valid when adhesion is included, at least as long as adhesion
is treated as a contact interaction. However, in this case the interfacial stiffness $dp_0/d\bar u$ must
be calculated including the adhesion (see Ref. \cite{CYang}). Plastic deformation can be included in
an approximate way as follows. If two solids are squeezed together at the pressure $p_0$ they will
deform elastically and, at short enough length scale, plastically. If the contact is now removed
the surfaces will be locally plastically deformed. Assume now that the surfaces
are moved into contact again at exactly the same position as the original contact, and with the same
squeezing pressure $p_0$ applied. In this case the solids will deform purely elastically and the
theory outlined in this paper can be (approximately) applied assuming that the surface roughness power spectrum
$\bar C(q)$ of the (plastically) deformed surface is known. In Ref. \cite{PSSR} we have described an approximately
way of how to obtain $\bar C(q)$ from $C(q)$ by defining (with $q=\zeta q_0$)
$$\bar C(q) = \left (1- {A_{\rm pl}(\zeta)\over A_{\rm pl}^0}\right )C(q)$$
where $A_{\rm pl}^0 = F_{\rm N}/\sigma_{\rm Y}$. The basic picture behind this definition is that
surface roughness at short length scales get smoothed out by plastic deformation, resulting in an
effective cut-off of the power spectrum for large wavevectors (corresponding to short distances).

\begin{figure}
\includegraphics[width=0.45\textwidth,angle=0]{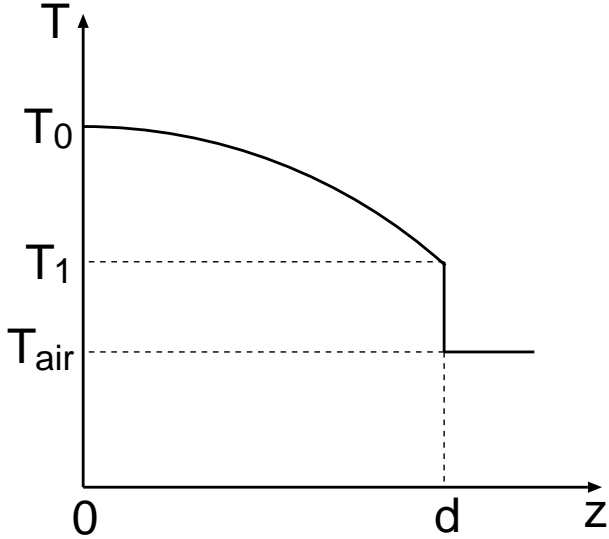}
\caption{\label{hest2}
Temperature distribution of rubber tread (thickness $d$) in contact with the air. The air temperature
(for $z > d$) and the temperature at the outer ($z=d$) and inner ($z=0$) rubber surfaces are denoted
by $T_{\rm air}$, $T_1$ and $T_0$, respectively.
}
\end{figure}

\vskip 0.3cm \textbf{6. Application to tires}

Here we will briefly discuss heat transfer in the context of tires. The rolling resistance
$\mu_{\rm R}$ of a tire determines the heat production in a tire during driving on a strait planar road
at a constant velocity $v$. In a stationary state the energy produced per unit time, $W=\mu_{\rm R} F_{\rm N} v$,
must equal the transfer of energy per unit time, from the tire to the surrounding atmosphere and to the road surface.
Here we will briefly discuss the relative importance of these two different contributions to the heat transfer.

Assume for simplicity that the frictional heat is produced uniformly in the tread rubber, and
assume a tire without tread pattern. Let $z$ be a coordinate axis perpendicular to the rubber surface.
In this case at stationary condition the temperature in the tread rubber satisfies
$T''(z)=-\dot q/\kappa$ where $\dot q$ is the frictional heat produced per unit volume and unit time. We assume that
the heat current vanish at the inner rubber surface ($z=0$, see Fig. \ref{hest2}),
so that $T'(0)=0$. Thus we get $T(z) = T_0-\dot q z^2 /2\kappa$.
The heat current at the outer rubber surface
$$J_0 =-\kappa T'(d) = \dot q d.\eqno(62)$$
The temperature of the outer surface of the tread rubber
$$T_1 = T(d) = T_0 -\dot q d^2 /2\kappa\eqno(63)$$
Let us now assume that the heat transfer to the surrounding
$$J_0 =\alpha (T_1-T_{\rm air})\eqno(64)$$
Combining (62)-(64) gives
$$T_1 = T_0-{T_0-T_{\rm air} \over 1+2\kappa/d\alpha}\eqno(65)$$
For rubber $\kappa \approx 0.2 \ {\rm W/mK}$ and with $d= 1 \ {\rm cm}$
and $\alpha \approx 100 \ {\rm W/m^2 K}$, as is typical for
(forced) convective heat transfer between a tire and (dry) air (see Appendix E and Ref. \cite{Oh}), we get
$$T_1 \approx 0.3 T_0 + 0.7 T_{\rm air}.$$
The temperature profile is shown (schematically) in Fig. \ref{hest2}.
In reality, the heat production, even during pure rolling, will
be somewhat larger close to the outer surface of the tread and the resulting temperature profile
in the tread rubber will therefore be more uniform than indicated by the analysis above.

Let us now discuss the relative importance of the contributions to the heat transfer
to the air and to the road.
We assume that the heat transfer to the atmosphere and to the road are proportional to the temperature
difference $T_1-T_{\rm air}$ and $T_1 - T_{\rm road}$, respectively.
We get
$$\mu_{\rm R} F_{\rm N} v = \alpha_{\rm air} A_{\rm surf} (T_1-T_{\rm air})
+ \alpha_{\rm road} A_0 (T_1 - T_{\rm road})\eqno(66)$$
where $A_{\rm surf}$ is the outer surface area of the tread,
and $A_0$ the nominal tire-road footprint area.
For rubber in contact with a road surface $\kappa$ in Eq. (22) is $ \approx 0.2  \ {\rm W/mK}$
and with $p_0/E^* \approx 0.04$ and $u_0 \approx 10^{-3} \ {\rm m}$ (as calculated for a typical case)
we get $\alpha_{\rm road} \approx 10 \ {\rm W/m^2 K}$ which is smaller than the contribution from
the forced convection. Since the nominal contact area between the tire and the road is much smaller than
the total rubber tread area, we conclude that the contribution from the area of real contact
between the road and the tire is rather unimportant. During fast acceleration wear process
may occur, involving the transfer of hot rubber particles to the road surface,
but such processes will not considered here. In addition, at the inlet of the tire-road footprint area,
air may be be compressed and then rapidly squeezed out
from the tire-road contact area resulting in strong forced convective cooling of the rubber surface
in the contact area. A similar process involving the inflow of air occur at the exit of the
tire-road footprint area.
A detailed study of this complex process is necessary in order to
accurately determine the
heat transfer from a tire to the surrounding atmosphere and the road surface.

For a passenger car tire
during driving on a strait planar road
at a constant velocity $v$,
the tire temperature which follows from (66) is in reasonably agreement with experiment. Thus,
using (66) we get
$$\Delta T =T_1-T_{\rm air} \approx {\mu_{\rm R} F_{\rm N} v \over \alpha_{\rm air} A_{\rm surf}}\eqno(67)$$
and with $\alpha_{\rm air} =  100 \ {\rm W/m^2 K}$, $A_{\rm surf} \approx 0.5 \ {\rm m^2}$
and $\mu_{\rm R} \approx 0.02$, $F_{\rm N} = 3500 \ {\rm N}$ and $v=30 \ {\rm m/s}$ we get
$\Delta T \approx 40 \ ^\circ {\rm C}$.

The discussion above has focused on the stationary state where the heat energy
produced in the tire per unit time is equal to the energy given off to the surrounding per unit time.
However, for a rolling tire
it may take a very long time to arrive at this stationary state. In the simplest picture, assuming
a uniform temperature in the tire rubber, we get from energy conservation
$$\rho C_{\rm v} {dT\over dt} = \dot q -{\alpha \over d} (T-T_{\rm air})$$
or, if $T(0)=T_{\rm air}$,
$$T(t) = T_{\rm air} +{\dot q d \over \alpha} \left (1-e^{-t/\tau}\right ),$$
where the relaxation time $\tau = \rho C_{\rm V} d /\alpha \approx 200 \ {\rm sec}$.
In reality, the temperature in the tire is not uniform,
and this will introduce another relaxation time $\tau'$, defined as the
time it takes for heat to diffuse a distance $d$, which is of order $\tau' = \rho C_{\rm V} d^2/\kappa$.
The ratio $\tau' / \tau  = \alpha d /\kappa$. For rubber $\kappa \approx 0.2 \ {\rm W/mK}$ and assuming
$d= 1 \ {\rm cm}$ and $\alpha = 100 \ {\rm W/m^2 K}$ gives $\tau' / \tau \approx 5$
or $\tau' \approx 10^3 \ {\rm sec}$. Experiment have shown that it typically takes $\sim 30 \ {\rm minutes}$
to fully build up the tire temperature during rolling\cite{Oh}.

Rubber friction depends sensitively on the temperature of the rubber, in particular the temperature close to the
rubber surface in contact with the road. The temperature in the surface region of a tire varies
rapidly in space and time, which must be considered when calculating the
rubber friction\cite{Flash}. The shortest time and length scales are related to the contact between the
road asperities and the rubber surface in the tire-road footprint contact area. During slip this generate
intense heating which varies over length scales from a few micrometer to several mm, and over time
scales shorter than the time a rubber patch stays in the footprint, which typically may be of order a few
milliseconds. During this short time very little heat is transferred to the surrounding, and very little
heat conduction has occurs inside the rubber, i.e., the heat energy mainly stays where it is produced by the
internal friction in the rubber. This result in a {\it flash temperature} effect,
which has a crucial influence on  rubber friction\cite{Flash}.
However, rubber friction also depends on the {\it background
temperature} (usually denoted by $T_0$), which varies relatively slowly in space and time, e.g.,
on time scales from the time $\sim 0.1 \ {\rm s}$ it
takes for the tire to perform a few rotations,
up to the time $\sim 30 \ {\rm minutes}$ necessary to build up the full tire temperature
after any change in the driving condition (e.g., from the start of driving).
Note that the time variation of the background temperature $T_0$
depends on the surrounding (e.g., the air and road temperatures, humidity, rain, ...) and on the driving history,
while the flash temperature effect mainly depends on the slip history of a tread block (or rubber surface patch)
in the footprint contact area, but not on the outside air or road temperature, or atmospheric condition.

\begin{figure}
\includegraphics[width=0.4\textwidth,angle=0]{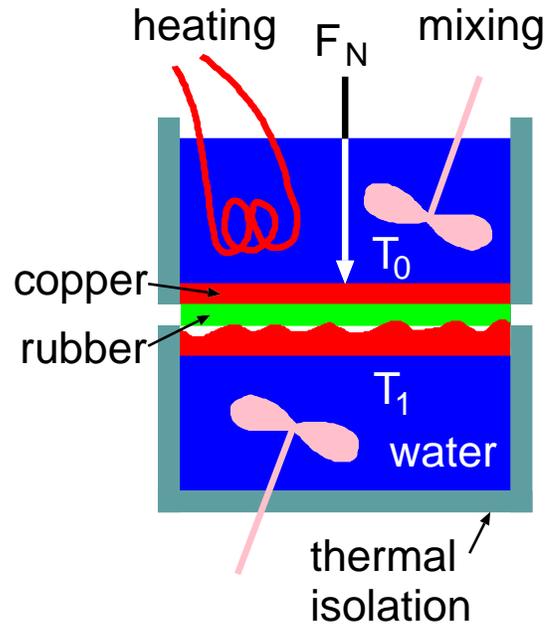}
\caption{\label{experiment}
Experiment to test the theory predictions for the heat transfer across interfaces.
The increase in the temperature $T_1(t)$ of the water in the lower container,
with increasing time $t$, determines the heat transfer
between the upper and lower water container.
}
\end{figure}

\vskip 0.3cm \textbf{7. A new experiment}

We have performed a very simple experiment to test the theoretical
predictions for the heat transfer.
The setup consists of two containers, both filled with distilled
water, standing on top of each other with a thin silicon rubber film
in between. The upper container is made from copper (inner diameter
$5 \ {\rm cm}$), 
and the water is heated to the boiling temperature
(i.e., $T_0=100 ^\circ\mathrm{C}$). The lower
container is made from PMMA with a cylindrical copper block at the top.
To study the effect of surface roughness on the heat transfer, 
the copper block can be replaced by 
another copper block with different surface roughness. In the experiments
presented below we used 3 copper blocks with different surface
roughness.

The temperature $T_1(t)$ of the water in the lower container will 
increase with time $t$ due to the heat
current $J_0$ flowing from the upper 
container to the lower container:
$$J_0 = \rho C_{\rm V} \dot T_1 d \eqno(68)$$
where $d$ is the height of the water column in the lower container
(in our experiment $d=3.5\ {\rm cm}$), and where $\rho$ and $C_{\rm V}$ are
the water mass density and heat capacity respectively. We measure
the temperature of the water in the lower container as a function of
time, starting at $25^\circ\mathrm{C}$. To obtain a uniform temperature
of the water in the lower container we mix it using 
a (magnetic-driven) rotating metal bar.

We have investigated the heat transfer using copper blocks with different surface roughness.
To prepare the rough surfaces, we have
pressed annealed (plastically soft) 
copper blocks with smooth surface against sandpaper, using
a hydraulic press. We repeated this procedure several times to
obtain randomly rough surfaces. The roughness of
the copper surfaces can be changed by using sandpaper of different grade 
(consisting of particles with different (average) diameter).
Due to the surface roughness, the
contact between the top surface of the lower container and the thin silicon rubber sheet
(thickness $d_0=2.5 \ {\rm mm}$) attached to the upper container, is only
partial. The bottom surface of the upper container has been highly
polished and we can neglect the heat resistance at this rubber-copper interface. 
Thus, most of the resistance to the heat flow arises from the heat diffusion through the rubber
sheet, and from the resistance to the heat flow at the interface
between the rubber and the rough copper block.

The rubber sheet (elastic modulus $E = 2.5 \ {\rm MPa}$, Poisson ration $\nu = 0.5$) was made from a
silicone elastomer (PDMS). We have used Polydimethylsiloxane because
of its almost purely elastic behavior on the time scales involved in our experiments. The PDMS
sample was prepared using a two-component kit (Sylgard 184)
purchased from Dow Corning (Midland, MI). This kit consists of a
base (vinyl-terminated polydimethylsiloxane) and a curing agent
(methylhydrosiloxane-dimethylsiloxane copolymer) with a suitable
catalyst. From these two components we prepared a mixture of 10:1
(base/cross linker) in weight. The mixture was degassed to remove
the trapped air induced by stirring from the mixing process and then
poured into cylindrical casts (diameter $5 \ \mathrm{cm}$ and height $d_0 = 2.5 \ \mathrm{%
mm}$). The bottom of these casts were made from glass to obtain
smooth surfaces (negligible roughness). The samples were cured in an
oven at $80 \ ^\circ\mathrm{C}$ for over 12 hours.

Using (3) we can write
$$J_0 \approx {T_0-T_1(t)\over d_0 \kappa_0^{-1}+\alpha^{-1}}\eqno(69)$$
where $\kappa_0$ the heat conductivity of the rubber.
Here we have neglected the influence of the copper blocks on the heat transfer resistance,
which is a good approximation because of the high thermal conductivity
of copper. Combining (68) and (69) gives
$$\tau \dot T_1 = T_0-T_1(t)$$
where the relaxation time
$$\tau_0 = \rho C_{\rm V} d \left ({d_0 \over \kappa_0}+{1\over \alpha} \right ).$$
If we assume that $\tau_0$ is time independent, we get
$$T_1(t) = T_0+\left [ T_1(0)-T_0 \right ] e^{-t/\tau_0}.\eqno(70a)$$

In the study above we have assumed that there is no heat transfer from the lower container to the surrounding.
However, if necessary one can easily take into account such a heat transfer:
If we assume that the heat transfer depends linearly on the 
temperature difference between the water and the surrounding
we can write
$$J_1 = \alpha_1 \left (T_1 - T_{\rm surr}\right)$$
In this case it is easy to show that (70a) is replaced with
$$T_1(t) = T_a+\left [ T_1(0)-T_a \right ] e^{-t/\tau}.\eqno(70b)$$
where $T_a$ is the temperature in the water after a long time (stationary state where $J_0=J_1$), 
and where the relaxation time $\tau$ now is given by
$$\tau = \rho C_{\rm V} d {T_a-T_{\rm surr} \over T_0 - T_{\rm surr}}
\left ({d_0 \over \kappa_0}+{1\over \alpha} \right ).$$

The heat transfer across the rubber--copper interface can occur
via the area of real contact, or via the non-contact area
via heat diffusion in the thin air film or via radiative heat transfer. Since all these heat
transfer processes act in parallel we have
$$\alpha \approx \alpha_{\rm gas} + \alpha_{\rm con} + \alpha_{\rm rad}.$$
Let us estimate the relative importance of these different contributions to $\alpha$.
Using the (diffusive) heat conductivity of air $\kappa_{\rm gas} \approx 0.02 \ {\rm W/mK}$
and assuming $\langle d^{-1} \rangle = (20 \ {\rm \mu m})^{-1}$ gives
$$ \alpha_{\rm gas} = \kappa_{\rm gas} \langle (d+\Lambda )^{-1}\rangle
\approx \kappa_{\rm gas} \langle d^{-1}\rangle \approx 1000 \ {\rm W/m^2K}.$$
Let us assume that
$p_0 \approx 0.01 \ {\rm MPa}$, $E^* \approx 2 \ {\rm MPa}$, $u_0 \approx 10 \ {\rm \mu m}$ and (for rubber)
$\kappa_0 = 0.2 \ {\rm W/mK}$. Thus
$$ \alpha_{\rm con} = {p_0 \kappa_0 \over E^* u_0} \approx 100 \ {\rm W/m^2K}.$$
Here we have used that $\kappa \approx \kappa_0$ (since the heat conductivity
$\kappa_1$ of copper is much higher than for the rubber).
Finally, assuming the radiative heat transfer is well approximated by the Stefan-Boltzmann law and
assuming that $(T_0-T_1)/T_1 << 1$, we get with $T_0 = 373 \ {\rm K}$
$$ \alpha_{\rm rad} \approx {\pi^2 k_{\rm B}^4 \over 60 \hbar^3 c^2} 4 T_0^3 \approx 10 \ {\rm W/m^2 K}$$

Note that $\alpha_{\rm rad}$ is independent of the squeezing pressure $p_0$, while $\alpha_{\rm con} \sim p_0$.
The pressure dependence of $\alpha_{\rm gas}$ will be discussed below.

In the experiment reported on below the silicon rubber film has the thickness $d_0 = 2.5 \ {\rm mm}$ so that
$d_0^{-1} \kappa_0 \approx 100 \ {\rm W/m^2K}$. Thus
$${1\over d_0^{-1} \kappa_0}+{1\over \alpha} \approx \left ( {1\over 100} + {1\over 1000+100+10}\right ) ({\rm W/m^2K})^{-1} $$
and it is clear from this equation that in the present case the thin rubber
film will give the dominant contribution to the heat resistance. This is in accordance with our experimental 
data presented below.


\begin{figure}
\includegraphics[width=0.45\textwidth,angle=0]{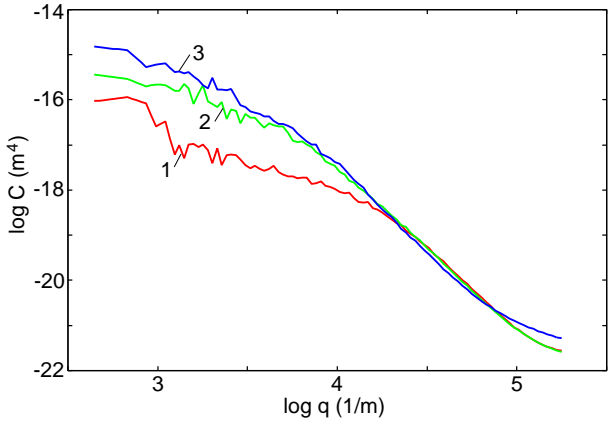}
\caption{\label{all3}
The surface roughness power spectrum of the three copper surfaces used in the experiment.
The surfaces {\bf 1}, {\bf 2} and {\bf 3} have the root-mean-square roughness
$42$, $88$ and $114 \ {\rm \mu m}$, respectively.
}
\end{figure}

\begin{figure}
\includegraphics[width=0.45\textwidth,angle=0]{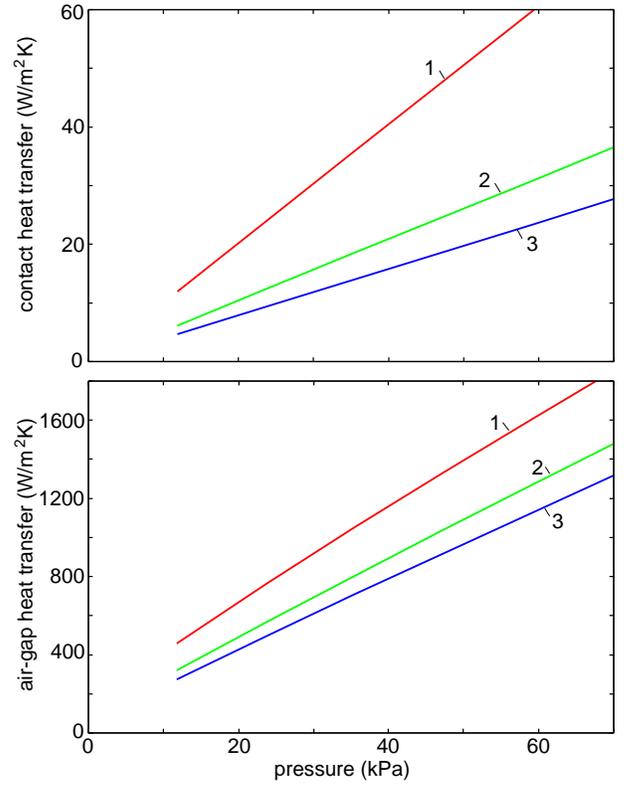}
\caption{\label{pressure.alphafluid.all3}
The variation of the of the heat transfer coefficient from the contact area ($\alpha_{\rm con}$)
and from the air-gap ($\alpha_{\rm gas}$) with the squeezing pressure.
The surfaces {\bf 1}, {\bf 2} and {\bf 3} 
have the power spectra's shown in Fig. \ref{all3}.
}
\end{figure}

\begin{figure}
\includegraphics[width=0.45\textwidth,angle=0]{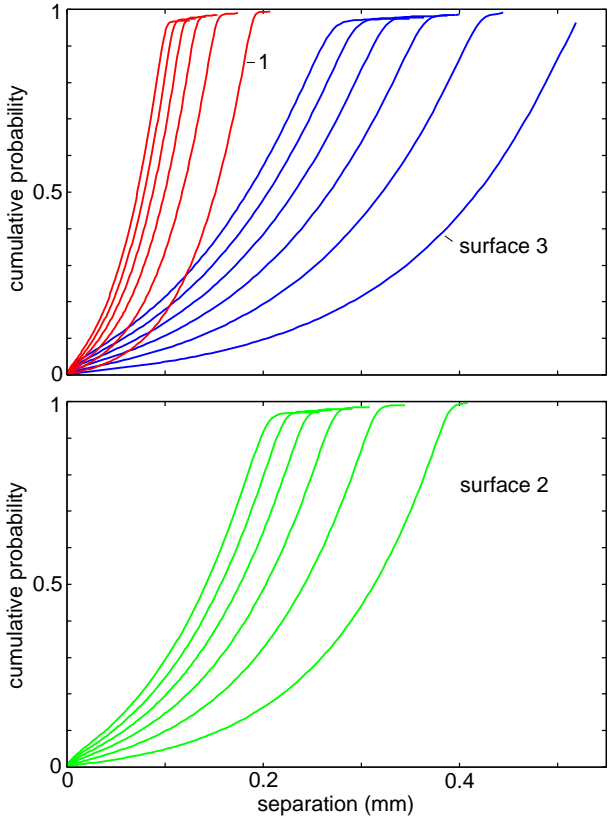}
\caption{\label{Surface1.and.3.cumulative.probability}
The variation of the cumulative probability with the height (or gap-separation) $u$.
The surfaces {\bf 1} and {\bf 3} (top) and {\bf 2} (bottom) 
have the power spectra's shown in Fig. \ref{all3}.
For each surface the curves are for the nominal squeezing pressures (from left to right):
$11.8$, $23.7$, $35.5$, $47.3$, $59.2$ and $71.0 \ {\rm kPa}$.
}
\end{figure}

\vskip 0.3cm \textbf{8. Experimental results and discussion}

To test the theory we have performed the experiment described in
Sec. 7. We have performed experiments on four different (copper)
substrate surfaces, namely one highly polished surface (surface {\bf 0}) with the 
root-mean-square (rms) roughness $64 \ {\rm nm}$, and for three rough
surfaces with the rms roughness $42$,
$88$ and $114 \ {\rm \mu m}$. In Fig. \ref{all3} we show the surface
roughness power spectrum of the three latter surfaces. Including only the
roughness with wavelength above $\sim 30 \ {\rm \mu m}$, the
rms slope of all three surfaces are of order
unity, and the normalized surface area $A/A_0 \approx 1.5$ in all
cases. 

In Fig. \ref{pressure.alphafluid.all3} we show for the surfaces {\bf 1}, {\bf 2} and {\bf 3},
the pressure dependence of heat transfer coefficient from the contact area ($\alpha_{\rm con}$)
and from the air-gap ($\alpha_{\rm gas}$). Note that both $\alpha_{\rm con}$ and $\alpha_{\rm gas}$ varies (nearly)
linearly with $p_0$. The latter may at first
appear remarkable because we know that at the low (nominal) squeezing pressures used
in the present calculation (where the area of real contact varies linearly
with $p_0$), the average surface separation $\bar u = \langle u \rangle$ depends logarithmically on $p_0$.
However, the heat transfer via heat diffusion in the air gap depends on $\langle (u+\Lambda )^{-1} \rangle$
which depends on $p_0$ almost linearly as long as $\bar u >> \Lambda$, which is obeyed in our case. 
This can be understood as follows: $\langle u \rangle$ is determined
mainly by the surface regions where the surface separation is close to its largest value. On the other
hand $\langle (u+\Lambda) ^{-1} \rangle$ is determined mainly by the surface regions where $u$ is very small, i.e.,
narrow strips (which we will refer to as boundary strips) 
of surface area close to the area of real contact. Now, for small $p_0$ the area of real contact
increases linearly with $p_0$ while the distribution of sizes of the contact
regions is independent of $p_0$. It follows that the total area of the boundary strips 
will also increase linearly with $p_0$. Thus, since $\langle (u+\Lambda )^{-1} \rangle$ is determined
mainly by this surface area, it follows that $\langle (u+ \Lambda )^{-1} \rangle$ will be nearly proportional to $p_0$.
We note that in Fig. \ref{HeatAlphaSiO2} $\alpha_{\rm gas}$ is nearly pressure independent, but this is due to the fact that
the (combined) surface in this case is extremely smooth 
(root-mean-square roughness $2.5 \ {\rm nm}$) so that the
$u$-term in $\langle (u+\Lambda)^{-1}\rangle$ can be neglected compared to the gas mean free path $\Lambda$,
giving a nearly pressure independent gas heat transfer coefficient. However, in the system studied
above $\bar u$ is much larger than $\Lambda$ and the result is nearly independent of $\Lambda$.

Note that in the present case (see Fig. 
\ref{pressure.alphafluid.all3})
$\alpha_{\rm gas} >> \alpha_{\rm con}$ so that the present experiment mainly test the theory for the heat
flow in the air gap. 

In Fig. \ref{Surface1.and.3.cumulative.probability}
we show the variation of the cumulative probability with the height (or gap-separation) $u$ for 
the surfaces {\bf 1} and {\bf 3} (top) and {\bf 3} (bottom).

\begin{figure}
\includegraphics[width=0.45\textwidth,angle=0]{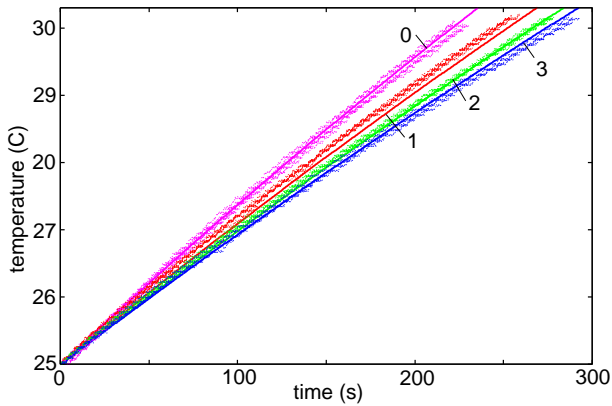}
\caption{\label{time.temp.surfaces.1.2.3.Boris}
The measured (dots) and calculated (solid lines) temperature in the lower container as a function of time. Results are for all four surfaces
and for the nominal squeezing pressure $p_0 = 0.012 \ {\rm MPa}$.
}
\end{figure}

\begin{figure}
\includegraphics[width=0.45\textwidth,angle=0]{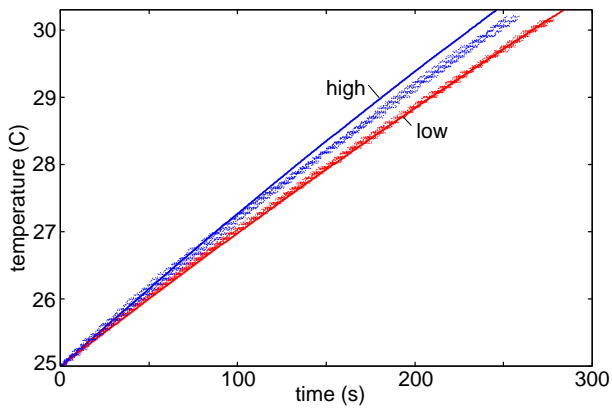}
\caption{\label{time.temp.surfaces.2.low.high.Boris}
The measured (dots) and calculated (solid lines) 
temperature in the lower container as a function of time. Results are for surface {\bf 2}
for the nominal squeezing pressure $p_0 = 0.012$ (lower curve) and $0.071 \ {\rm MPa}$ (upper curve).
}
\end{figure}

In Fig. \ref{time.temp.surfaces.1.2.3.Boris} we show
the measured (dots) and calculated [using (70b)] (solid lines) temperature in the 
lower container as a function of time. Results are for all four surfaces
and for the nominal squeezing pressure $p_0 = 0.012 \ {\rm MPa}$.
In Fig. \ref{time.temp.surfaces.2.low.high.Boris} we show
the measured (dots) and calculated (solid lines) 
temperature in the lower container as a function of time. Results are for surface {\bf 2}
for the nominal squeezing pressure $p_0 = 0.012$ (lower curve) and $0.071 \ {\rm MPa}$ (upper curve).
Note that there is no fitting parameter in the theory calculations, and the agreement between theory and
experiment is relative good.

The heat resistance of the system studied above is dominated by the thin rubber film. The reason for this is
the low heat conductivity of rubber (roughly 100 times lower than for metals). For direct metal-metal contact
the contact resistance will be much more important. However, for very rough surfaces it is likely that plastic
flow is observed already at such low magnification (corresponding to large length scales) that it will affect
the contact resistance. Nevertheless, it is interesting to compare the theory predictions for elastic contact with
experimental data for metal-metal contacts.

\begin{figure}
\includegraphics[width=0.45\textwidth,angle=0]{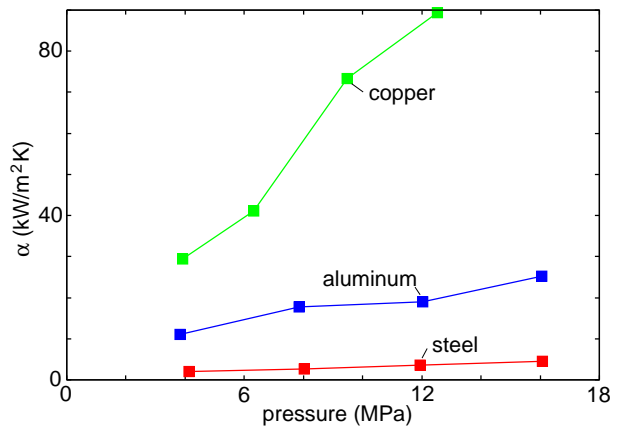}
\caption{\label{Fe.Cu.Al.new}
Variation of the heat transfer coefficient $\alpha$ with the squeezing pressure $p_0$
for metal-metal contact with steel, copper and aluminum. The surfaces have the effective (or combined)
root-mean-square surface roughness values $h_{\rm rms}=7.2 \ {\rm \mu m}$ (steel), $2.2 \ {\rm \mu m}$ (copper)
and $5.0 \ {\rm \mu m}$ (aluminum). The heat conductivity of the metals are
$\kappa = 54 \ {\rm W/mK}$ (steel), $381 \ {\rm W/mK}$ (copper) and $174 \ {\rm W/mK}$ (aluminum).
Based on experimental data from Ref. \cite{data1}.
}
\end{figure}

In Fig. \ref{Fe.Cu.Al.new} we show the measured heat transfer coefficient for metal-metal contacts
with steel, copper and aluminum\cite{data1}. The surfaces have the effective (or combined)
rms surface roughness $h_{\rm rms}=7.2 \ {\rm \mu m}$ (steel), $2.2 \ {\rm \mu m}$ (Cu)
and $5.0 \ {\rm \mu m}$ (Al). Assume that the variation of $\alpha$ with $p_0$ is mainly due to
the area of real contact, i.e., we neglect the heat transfer via the thin air
film between the surfaces. Fitting the data points in Fig.
\ref{Fe.Cu.Al.new} with strait lines gives the slope
${d\alpha / dp_0} ({\rm exp})$ (in units of ${\rm {m/sK}}$):
$$2\times 10^{-4} \ ({\rm steel}), \ \ \ \ \ 7 \times 10^{-3} \ ({\rm Cu}), \ \ \ \ \ 1.2 \times 10^{-3} \ ({\rm Al})$$
Using (26a) with $u_0 \approx 0.4 h_{\rm rms}$ (here we have assumed $\gamma = 0.4$) gives ${d\alpha / dp_0} ({\rm theory})= \kappa/E^*u_0$:
$$1\times 10^{-4} \ ({\rm steel}), \ \ \ \ \ 4 \times 10^{-3} \ ({\rm Cu}), \ \ \ \ \ 1.3 \times 10^{-3}  \ ({\rm Al})$$
The agreement between theory and experiment is very good taking into account that plastic deformation may
have some influence on the result, and that an accurate analysis requires the full
surface roughness power spectrum $C(q)$ (in order to calculate $u_0$ accurately, and in order to
include plastic deformation if necessary (see Sec. 5)),
which was not reported on in Ref. \cite{data1}.
We note that experimental results such as those presented in Fig. \ref{Fe.Cu.Al.new} are usually analyzed with
a phenomenological model which assumes plastic flow and neglect elastic deformation. In this theory
the heat transfer coefficient\cite{Yovanovich}
$$\alpha \approx {\kappa s p_0\over h_{\rm rms}\sigma_{\rm Y}}\eqno(71)$$
is proportional to the rms {\it surface slope} $s$, but it is well
known that this quantity is dominated by the very shortest wavelength roughness which in fact makes the theory ill-defined.
In Ref. \cite{data1} the data presented in Fig. \ref{Fe.Cu.Al.new} was analyzed using (71) with $s=0.035$, $0.006$
and $0.03$ for the steel, Cu and Au surfaces, respectively.
However, analysis of polished surfaces with similar rms
roughness as used in the experiments usually gives slopes of order unity when all roughness down to the nanometer
is included in the analysis\cite{unpublished}.
Using $s\approx 1$ in (71) gives heat transfer coefficients roughly $\sim 100$ times
larger than observed in the experiments. (In our theory [Eq. (26a)]
$s/\sigma_{\rm Y}$ in (71) is replaced by $1/E^*$, and since
typically $E^*/\sigma_Y \approx 100$, our theory
is consistent with experimental observations.)\cite{argued}
We conclude that the theory behind (71) is incorrect or incomplete. A theory which
includes both elastic and plastic deformation was described in Sec. 5.

\vskip 0.3cm \textbf{9. Electric contact resistance}

It is easy to show that the problem of the electrical contact resistance is mathematically
equivalent to the problem of the thermal contact resistance. Thus, the
electric current (per unit nominal contact area) $J_0$ through an interface
between solids with randomly rough surfaces can be related to the electric potential drop
$\Delta \phi$ at the interface via $J_0 =\alpha' \Delta \phi$ where, in analogy with
(25),
$$\alpha' = {p_0 \kappa' \over E^* u_0}\eqno(72)$$
where $\kappa'$ is the electrical conductivity. However, from a practical point of view
the problem of the electrical contact resistance is more complex than for the heat contact
resistance because of the great sensitivity of the electric conductivity on the type
of material (see Appendix D). Thus, in a metal-metal contact the contact resistance will depend sensitively
on if the thin insulating oxide layers, which covers most metals, are fractured, so that
direct metal-metal contact can occur. On the other hand, in most cases there will be a negligible
contribution to the electric conductivity from the non-contact regions.

\vskip 0.3cm \textbf{10. Summary and conclusion}

We have studied the heat transfer between elastic solids with randomly rough but nominally flat surfaces
squeezed in contact with the pressure $p_0$. Our approach is based on studying the heat flow and contact mechanics in
wavevector space rather than real space which has the advantage that we do not need to consider the
very complex fractal-like shape of the contact regions in real space. We have included both the heat flow in the
area of real contact as well as the heat flow across the non-contact surface region. For the latter contribution we have
included the heat transfer both from the fluctuating electromagnetic field (which surrounds all material objects),
and the heat flow via the surrounding gas or liquid. We have also studied the contribution to the heat transfer from capillary bridges, which
form spontaneously in a humid atmosphere (e.g., as a result of organic and water contamination
films which occur on most solid surfaces in the normal atmosphere).
We have presented an illustrative application relevant for MEMS applications involving very
smooth amorphous silicon dioxide surfaces. In this case we find that all the mentioned heat transfer processes
may be roughly of equal importance.

We have briefly discussed the role of plastic deformation and adhesion on the contact heat resistance. We have pointed out that
even if plastic deformation and adhesion are important at short length scale (or high magnification) they may have a negligible
influence on the heat transfer since the elastic energy stored in the asperity contact regions, which mainly determines both the
interfacial separation and the contact heat transfer coefficient, is usually mainly determined by the
long-wavelength surface roughness components, at least for fractal-like surfaces with fractal dimension $D_{\rm f} < 2.5$
(which is typically obeyed for natural surfaces and surfaces of engineering interest).

\vskip 0.5cm
{\bf Acknowledgments}

We thank Christian Schulze (ISAC, RWTH Aachen University) for help
with the measurement of the surface topography of the copper
surfaces. A.I.V. acknowledges financial support from the Russian
Foundation for Basic Research (Grant N 08-02-00141-a) and from DFG.
This work, as part of the European Science Foundation EUROCORES
Program FANAS, was supported from funds by the DFG and the EC Sixth
Framework Program, under contract N ERAS-CT-2003-980409.

\vskip 1cm

{\bf Appendix A}

In Sec. 2.2.1 we have assumed that
$${1\over J_0^2}\int d^2 q {1\over q} \langle |\Delta J_z({\bf q})|^2 \rangle \approx
{1\over p_0^2}\int d^2 q {1\over q} \langle |\Delta \sigma_z({\bf q})|^2 \rangle \eqno(A1)$$
This equation is a consequence of the fact that
for elastic solids with randomly rough surfaces the heat transfer coefficient depends mainly on the
geometrical distribution of the contact area. This can be understood as follows.
Let ${\rm x}_n$ denote the center of the contact spot $n$ and let $I_n$ be the heat current through
the same contact spot. We now approximate
$$J_z({\bf x}) \approx \sum_n I_n \delta ({\bf x}-{\bf x}_n).$$
Thus
$$A_0 J_0 = \sum_n I_n$$
and
$$J_z({\bf q}) = {1\over (2\pi )^2} \sum_n I_n e^{-i{\bf q}\cdot {\bf x}_n}$$
Thus the left hand side (LHS) of (A1) becomes
$${\rm LHS}\approx \left ({A_0\over (2\pi )^2}\right )^2 \left (\sum_n I_n \right )^{-2}$$
$$\times \sum^\prime_{mn} I_n I_m \int d^2q \
{1\over q} e^{i{\bf q}\cdot ({\bf x}_m-{\bf x}_n)}\eqno(A2)$$
where the prime on the summation indicate that the term $m=n$ is excluded from the sum.
Next note that
$$\int d^2q \ {1\over q} e^{i{\bf q}\cdot ({\bf x}_m-{\bf x}_n)}= {4 \pi
\over |{\bf x}_m-{\bf x}_n |} \eqno(A3)$$
Substituting (A3) in (A2) gives
$${\rm LHS} \approx {A_0^2\over 4 \pi^3} \left (\sum_n I_n \right )^{-2}
\sum^\prime_{mn} {I_m I_n \over |{\bf x}_m-{\bf x}_n|} \eqno(A4)$$
If one assume that there is no correlation between the magnitude of $I_n$
(determined by the size of the contact)
and its position, we can replace the individual current $I_n$ in
the double summation in (A2) by their mean and get
$${\rm LHS}
\approx  {1\over 4 n^2 \pi^3}
\sum^\prime_{mn} {1 \over |{\bf x}_m-{\bf x}_n |}\eqno(A5)$$
where $n=N/A_0$ is the concentration of contact spots and $N$ the total number of contact spots.

In the same way as above one can simplify the expression involving the
normal stress (right hand side (RHS) of (A1)). We write
$$\sigma({\bf x}) = \sum_n f_n \delta ({\bf x}-{\bf x}_n)$$
where $f_n$ is the normal force acting in the contact $n$.
Using this equation the RHS of (A1) becomes
$${\rm RHS} \approx {A_0^2\over 4 \pi^3} \left (\sum_n f_n \right )^{-2}
\sum^\prime_{mn} {f_m f_n \over |{\bf x}_m-{\bf x}_n|}\eqno(A6)$$
If one assume that there is no correlation between the magnitude of $f_n$
and its position, we can replace the individual current $f_n$
in the double summation in (A6) by their mean and get
$${\rm RHS} \approx  {1\over 4 n^2 \pi^3}
\sum^\prime_{mn} {1 \over |{\bf x}_m-{\bf x}_n |}\eqno(A7)$$
Thus, ${\rm LHS} \approx {\rm RHS}$ and we have proved the (approximate) equality (A1).

Substituting (A5) in (20) gives
$${1\over \alpha} \approx {1 \over \pi \kappa n} {1\over N} \sum^\prime_{mn}
{1 \over |{\bf x}_m-{\bf x}_n|}\eqno(A8)$$
which agree with the derivation of Greenwood\cite{GreenW}. We refer to the article of Greenwood for an interesting discussion
about the contact resistance based on the (approximate) expression (A8) for the contact resistance.

\vskip 0.5cm

{\bf Appendix B}

The normal (interfacial) stress $\sigma_z({\bf x})$ and the difference in the
surface displacement $u_{0z}({\bf x})-u_{1z}({\bf x})$ at the interface
can be considered to depend on the average interfacial separation $\bar u$.
The derivatives of these quantities with respect to $\bar u$ are denoted by $\sigma'_z$ and
$\phi$. In Appendix C we show that
$$\phi({\bf q}) = \delta ({\bf q}) - {2\over E^* q} \Delta \sigma'_z({\bf q}).\eqno(B1)$$
Note that (15) and (B1) are very similar. Thus, if we multiply both sides
of (B1) with $M$ and define $M\phi = \psi$ then (B1) takes the form
$$\psi ({\bf q}) = M \delta ({\bf q})-{\mu \over \kappa q} \Delta \sigma'_z({\bf q})\eqno(B2)$$
where
$$\mu = {2 M \kappa \over E^*}\eqno(B3)$$
Eq. (B2) is identical to (15) if we write
$$J_z({\bf q}) = \mu \sigma'_z({\bf q}),\eqno(B4)$$
or, equivalently,
$$J_z({\bf x})/J_0 =\sigma'_z ({\bf x})/p'_0$$
where $p'_0$ is the normal stiffness. We note that (B4) implies that the current density
$J_z({\bf x})$ will be non-vanishing
exactly where the normal stress $\sigma_z ({\bf x})$ is non-vanishing, which must be
obeyed in the present case, where all the heat current flow through the area of real contact.
However, in order for $J_z({\bf x})$ to be proportional to $\sigma'_z({\bf x})$ it is not enough that these functions
obey similar (in the sense discussed above) differential equations, but
both problems must also involve similar
boundary conditions. Now in the area of non-contact both $J_z$ and $\sigma_z$ and hence $\sigma'_z$ must vanish. In the area of
real contact the temperature field $T$ is continuous so that $\psi = T({\bf x},-0)-T({\bf x},+0) = 0$,
while the displacement field satisfies $\Phi = u_{0z}-u_{1z} = h({\bf x})$
so that (since $h({\bf x})$ is independent of $\bar u$), $\phi = 0$ in the area of real contact. Thus,
both problems involves the same boundary conditions and $J_z$ and $\sigma'_z$ must
therefore be proportional to each other.

Note that (B4) gives $J_0 = \mu p'_0$. Substituting (B3) in this equation and using the definition (16)
gives an equation of the form (3) with
$$\alpha = - {\kappa \over E^*} {dp_0 \over d\bar u }.$$
This {\it exact} relation between the heat transfer coefficient and the normal stiffness per unit area
has already been derived by Barber\cite{Barber} using a someone different approach.

\vskip 0.5cm
{\bf Appendix C}

In Ref. \cite{JCPpers}  it was shown that the normal displacement $u_{0z}$ is related to the
normal stress $\sigma_z$ via
$$u_{0z}({\bf q})= -{2\over E_0^* q}\sigma_z({\bf q}),\eqno(C1)$$
where $E_0^* = E_0/(1-\nu_0^2)$.
In a similar way
$$u_{1z}({\bf q}) = {2\over E_1^* q} \sigma_z({\bf q}).\eqno(C2)$$
Let $\Phi = u_{0z}-u_{1z}$ be the difference between the (interfacial) surface displacement fields.
Using (C1) and (C2) gives
$$\Phi({\bf q}) = -{2\over E^* q} \sigma_z({\bf q})\eqno(C3)$$
where
$${1\over E^*} = {1\over E_0^*}+{1\over E_1^*}$$
Note that the average of $\Phi({\bf x})$ is the average separation between the surfaces which we
denote by $\bar u$. Thus if
$$\sigma_z({\bf x}) = p_0 + \Delta \sigma_z({\bf x})$$
we get
$$\Phi({\bf q}) = \bar u \delta({\bf q})-{2\over E^* q} \Delta \sigma_z({\bf q})\eqno(C4)$$
As the squeezing pressure $p_0$ increases, the average separation $\bar u$ will decrease and we can
consider $p_0$ as a function of $\bar u$. The quantity $p'_0(\bar u)$ is referred to as the normal
stiffness per unit nominal contact area. Taking the derivative of (C4) with respect to $\bar u$ gives
$$\phi({\bf q}) = \delta({\bf q})-{2\over E^* q} \Delta \sigma'_z({\bf q})\eqno(C5)$$
where $\sigma'_z$ is the derivative of $\sigma_z$ with respect to $\bar u$ and where
$\phi = \Phi'$ is the derivative of $\Phi$ with respect to $\bar u$.


\vskip 0.5cm
{\bf Appendix D}

Heat conduction
result from the collisions between atoms
as in fluids, or by free electron diffusion
as predominant in metals, or phonon diffusion as predominant in insulators.
In liquids and gases,
the molecules are usually further apart than in solids, giving a lower chance of molecules colliding and passing on thermal energy.
Metals are usually the best conductors of thermal energy. This is due to the free-moving
electrons which are able to transfer thermal energy rapidly through the metal. However, the difference in the thermal conductivity of metals and non-metals
are usually not more than a factor $\sim 100$.
Typical values for the heat conductivity are
$\kappa \approx 100 \ {\rm W/mK}$ for metals, $\approx 1 \ {\rm W/mK}$ for insulators (e.g., metal oxides or polymers),
$\approx 0.1 \ {\rm W/mK}$ for fluids (but for water $\kappa \approx 0.6 \ {\rm W/mK}$) and $\approx 0.02 \ {\rm W/mK}$
for gases at normal atmospheric pressure and room temperature.

In contrast to thermal heat transfer, electric conduction always involves the motion of charged particles (electrons or ions).
For this reason the electric contact resistance is much more sensitive to oxide or contamination
layers at the contacting interface then for the heat transfer.
For the electric conduction the variation of the conductivity between good
conductors (most metals), with the typical electric conductivity $\kappa' \approx 10^7 \ {\rm (\Omega m)^{-1}}$, and bad conductors
such as silicon dioxide glass or (natural) rubber where $\kappa' \approx 10^{-14} \ {\rm (\Omega m)^{-1}}$, is huge.
This makes the electrical contact resistance of metals sensitive to (nanometer) thin oxide or contamination layers.
However, as pointed out in the Introduction, if there is a large number of small breaks in the film, the resistance may
be almost as low as with no film.

\vskip 0.5cm
{\bf Appendix E}

Here we briefly summarize some results related to forced convective heat transfer\cite{Landau}.
When a fluid (e.g., air) flow around a solid object the tangential (and the normal) component
of the fluid velocity usually vanish on the surface of the solid. This result in the formation of a
thin boundary layer (thickness $\delta$) at the surface of the solid where the fluid velocity rapidly increases from
zero to some value which is of order the main stream velocity outside of the solid. If the
temperature $T_1$ at the solid surface is different from the fluid temperature $T_{\rm fluid}$,
the fluid temperature in the boundary layer will also change rapidly from $T_1$ to $T_{\rm fluid}$.
Depending on the fluid flow velocity, the fluid viscosity and the dimension of the solid object
the flow will be laminar or turbulent, and the heat transfer process is fundamentally
different in these two limiting cases. In a typical case (for air) the thickness $\delta \approx 1 \ {\rm mm}$
and the heat transfer coefficient $\alpha \approx \kappa / \delta \approx 10 \ {\rm W/m^2K}$.

Let us consider heat transfer from a rotating disk as a model for the heat transfer
from a tire\cite{Allen}.
In this case it has been shown\cite{Popiel} that
fully turbulent flow occur if the Reynolds number ${\rm Re} > 2.5\times 10^5$ where
$${\rm Re}= {\omega R^2\over \nu} = {v_{\rm R} R \over \nu}$$
where $R$ is the radius of the disk (or rather the distance from the center of
the disk to some surface patch on the disk), $\omega$ the angular velocity and $\nu$ the kinematic
viscosity of air. In typical tire applications ${\rm Re} > 2.5\times 10^5$ so turbulent flow will prevail
in most tire applications. In this case the heat transfer coefficient is given approximately by\cite{Popiel}:
$$\alpha_{\rm air} \approx 0.019 {\kappa_{\rm air} \over R} \left ( {v_{\rm R} R \over \nu} \right )^{0.8}.$$
As an example, at $T=300 \ {\rm K}$ for air $\nu = 15.7 \times 10^{-6} \ {\rm m^2/s}$
and $\kappa_{\rm air} = 0.025 \ {\rm W/mK}$
and assuming
$R=0.3 \ {\rm m}$ and $v_{\rm R} = 30 \ {\rm m/s}$ we get $\alpha_{\rm air} \approx 63 \ {\rm W/m^2K}$.


\begin{thebibliography}{99}
\bibitem{unpublished}
As an example, using AFM we have measured the height profile of a polished steel surface over a $10 {\rm \mu m}\times 10 {\rm \mu m}$
surface area with the resolution $a=20 \ {\rm nm}$. From the numerical data we calculated
the root-mean-square (rms) roughness $h_{\rm rms} \approx 0.1 \ {\rm \mu m}$ and the rms
slope $s \approx 0.6$. Increasing the lateral resolution would increase the slope further since the slope is mainly determined by the
short wavelength roughness.
\bibitem{argued}
It may be argued that, due to plastic deformation, the slope in (71) should be calculated
including only the roughness with wavelength above some cut-off length. However, no discussion of this point was
presented in Ref. \cite{Yovanovich}.
\end{thebibliography}
\end{document}